\pdfoutput=1
\documentclass[aps,prd,amsmath,floats,floatfix,twocolumn,
superscriptaddress,nofootinbib,showpacs,longbibliography]{revtex4-1}

\usepackage{packages}

\begin{document}

\title{
Searching for ringdown higher modes with a numerical relativity-informed\\ post-merger model
}

\newcommand{\LdIT}{\affiliation{Laboratoire des 2 Infinis - Toulouse (L2IT-IN2P3), Université de Toulouse, CNRS, UPS, F-31062 Toulouse Cedex 9, France}}
\newcommand{\Pisa}{\affiliation{Dipartimento di Fisica “Enrico Fermi”, Università di Pisa, Pisa I-56127, Italy}}
\newcommand{\INFN}{\affiliation{INFN, Sezione di Pisa, I-56127 Pisa, Italy}}
\newcommand{\NBI}{\affiliation{Niels Bohr International Academy, Niels Bohr Institute, Blegdamsvej 17, 2100 Copenhagen, Denmark}}
\newcommand{\Jena}{\affiliation{Theoretisch-Physikalisches Institut, Friedrich-Schiller-Universit{\"a}t Jena, 07743, Jena, Germany}}

\author{Vasco Gennari}
\email{vasco.gennari@l2it.in2p3.fr}
\LdIT\Pisa\INFN\NBI

\author{Gregorio Carullo}
%\email{gregorio.carullo@unipi.it}
\Pisa\INFN\Jena\NBI

\author{Walter Del Pozzo}
%\email{walter.delpozzo@unipi.it}
\Pisa\INFN

% Because hyperref only gets the *last* author, we need to be explicit.
\hypersetup{pdfauthor={Gennari et al.}}

\date{\today}

%==========================================================================
\begin{abstract}

Robust measurements of multiple black hole vibrational modes provide a unique opportunity to characterise gravity in extreme curvature and dynamical regimes, to better investigate the nature of compact objects and search for signs of new physics. We use a numerically-tuned quasicircular non-precessing ringdown model, \texttt{TEOBPM}, and the \texttt{pyRing} analysis infrastructure to perform a time-domain spectroscopic analysis of the third catalog of transient gravitational-wave signals, GWTC-3, searching for higher angular modes. The \texttt{TEOBPM} model effectively includes non-linearities in the early post-merger signal portion, and carries information about the progenitors parameters through time-dependent excitation amplitudes of the black hole quasinormal modes. Such a strategy allows us to accurately model the full post-merger emission, recovering higher signal-to-noise ratios compared to templates based on more agnostic superpositions of damped-sinusoids. We find weak evidence for the presence of $(l,m)=(3,3)$ [$(l,m)=(2,1)$] mode in several events, with the largest Bayes factor in favour of this mode being $\mathcal{B} \simeq 2.6$ [$\mathcal{B} \simeq 1.2$] within the support of the peak time distribution. For GW190521, we observe $ \mathcal{B} \simeq 5.1$, but only for times outside the peak time support reconstructed using the highly accurate \texttt{NRSur7dq4} model, indicating significant systematics affecting such putative detection. Allowing for deviations from general relativity under the assumption of the presence of two modes, we find tentative support for the Kerr ``final state conjecture''. Our work showcases a systematic methodology to robustly identify and characterise higher angular modes in ringdown-only signals, highlighting the significant impact of modelling assumptions and peak time uncertainty on spectroscopic measurements, at current signal-to-noise ratios.

\end{abstract}

\maketitle
\tableofcontents

%==========================================================================
\section{Introduction}
\label{sec:introduction}

% RD: importance, higher modes, tests of no-hair
Observations of gravitational waves (GWs) from compact binary coalescences through the LIGO-Virgo-KAGRA interferometric network~\cite{AdvLIGO, AdvVirgo,PhysRevD.88.043007} are revolutionising our understanding of the Universe, from black hole (BH) physics to astrophysics, cosmology and fundamental physics~\cite{LIGOScientific:2021djp, LIGOScientific:2021sio, KAGRA:2021duu, LIGOScientific:2021aug}.
As GWs from the inspiral and plunge phases carry information about the long-range dynamics of two coalescing BHs, the merger and \textit{ringdown} (RD) regimes provide unique access to extreme curvatures and strong fields, being arguably the best route to investigate the nature of BHs and challenge our current understanding of gravity~\cite{LIGOScientific:2016lio, LIGOScientific:2019fpa, LIGOScientific:2020tif, LIGOScientific:2021sio,Berti:2015itd,Berti:2018vdi,Cardoso:2019rvt}.

The linear theory of BH perturbations~\cite{Regge:1957td,Zerilli:1970wzz,Zerilli:1970se,Teukolsky:1972my,Teukolsky:1973ha,Press:1973zz,Teukolsky:1974yv} predicts that
that the main RD contribution consists of a sum of spacetime vibrations known as \textit{quasinormal modes} (QNMs), which are exponentially damped harmonic oscillations~\cite{Vishveshwara:1970zz,Chandrasekhar:1975zza, Nollert:1999, Ferrari:2007dd, Berti:2009kk}.
The excitation of different modes depends on the specific process that perturbs the BH, and for remnants from quasicircular binary coalescences only one long-lived mode typically dominates the RD emission.
Asymmetries in the system are responsible for the excitation of subdominant \textit{higher angular modes} (HMs)~\cite{Berti:2007fi,Kamaretsos:2011um, Kamaretsos:2012bs,London:2014cma, London:2018gaq, Bhagwat:2019bwv, Bhagwat:2019dtm, Mills:2020thr, JimenezForteza:2020cve, Ota:2021ypb, Cheung:2023vki}.
The measurement of the frequencies and damping times of these HMs allows to directly question the BH paradigm by testing the Kerr ``final state conjecture''~\footnote{The no-hair conjecture, plus the conjecture that the Kerr solution is a dynamical attractor for BHs relaxing in realistic astrophysical scenarios, sometimes more loosely referred to just as ``testing the no-hair theorem''.}~\cite{Loutrel:2020wbw,Penrose1982SOMEUP,Ginzburg,Zeldovich,Zeldovich2,Israel,Carter,Hawking1972,Robinson,Bunting, Mazur}. 
However, the observation of HMs in ringdown signals is extremely challenging at current detectors sensitivities, and their characterisation is further complicated by several delicate issues affecting RD analyses.

% Nonlinearities and starting time, EOB models
In fact, current studies based on QNM superpositions rely on the assumption that, at late enough times during the postmerger relaxation process, the spacetime is well-described by linear perturbation theory.
This assumption manifests itself in, e.g., a constant final mass and spin, and of (asymptotically) constant ringdown complex amplitudes\footnote{Even linear perturbation theory predicts an amplitude growth~\cite{Berti:2006wq, Lagos:2022otp,Andersson:1996cm,Szpak:2004sf}, but such growth is dependent on the initial data, and not easy to disentangle from other nonlinear effects in a binary merger. Lacking a detailed understanding of this process, amplitudes are often assumed to be constant at late enough times.}.
Indeed, the merger and early postmerger phases are known to contain nonlinear features~\cite{Gleiser:1996yc}, and a sufficient amount of time needs to pass before the linear regime starts to hold and the above assumptions are justified~\cite{Buonanno:2006ui,London:2014cma,Carullo:2018sfu, Bhagwat:2017tkm, London:2018gaq}.
Such complications are among the motivations behind the large amount of recent progress in modelling higher-order effects, which might soon help to extend ringdown models for comparable-mass binaries closer to the signal peak~\cite{Sberna:2021eui,Mitman:2022qdl, Lagos:2022otp, Cheung:2022rbm,Bucciotti:2023ets, Perrone:2023jzq, Redondo-Yuste:2023seq}.
Although techniques to estimate the (signal-to-noise ratio dependent) start of the linear regime validity have been developed on both numerical relativity (NR) simulations~\cite{Bhagwat:2017tkm} and real data~\cite{Carullo:2019flw, Finch:2022ynt}, the applicability time-window of QNM superpositions remains the most relevant systematic in most RD analyses.

To address this problem, immediately after the first binary numerical evolutions~\cite{Pretorius:2005gq,Campanelli:2005dd,Baker:2005vv}, RD models going beyond QNM superpositions have been developed, e.g.~\cite{Baker:2008mj}.
The time-dependence of the BH parameters and amplitudes is now phenomenologically included through a flexible function with several free parameters, thus relaxing the requirement to describe the process from first principles.
In particular, such models are required to complete effective-one-body (EOB)~\cite{Buonanno:1998gg, Buonanno:2000ef, Damour:2001tu} waveforms beyond merger, and in Refs.~\cite{Damour:2014yha, DelPozzo:2016kmd, Bohe:2016gbl, Cotesta:2018fcv,Nagar:2019wds,Nagar:2020pcj, Pompili:2023tna} accurate templates capable of modelling the entire post-merger emission were presented. 
Such models allow to include larger portions of the GW signal compared to a superposition of QNMs with constant amplitudes, increasing the available signal-to-noise ratio (SNR).
Although they are less sensitive to certain exotic deviations from a Kerr relaxation, by including more information and being less agnostic about the physics involved, they allow to extract a larger amount of information from the data.
As extensively discussed in Ref.~\cite{Baibhav:2023clw}, accounting for such nonlinear contributions is of paramount importance.
Failing to do so leads to overfitting, misinterpreting the physical QNM content of the signal, and in turn to faulty QNM ``detections'' plagued by systematic uncertainties caused by the inappropriate assumption of constant amplitude models close to the signal peak.

% TEOBPM analysis of GWTC-3
In this work, we rely on an accurate phenomenological model covering the entire post-peak emission, \texttt{TEOBPM}~\cite{Damour:2014yha, DelPozzo:2016kmd,Nagar:2019wds,Nagar:2020pcj}, and use it to perform a comprehensive Bayesian analysis of the third catalog of GW events, GWTC-3~\cite{LIGOScientific:2021djp}, using \texttt{pyRing}~\cite{pyRing}, a \texttt{python} software tailored to perform a time-domain Bayesian analysis on BH ringdown signals.
Our analysis allows to systematically assess the detectability of HMs in the RD at current sensitivities, extending previous LVK ringdown-only studies which relied on models based on pure QNM superpositions~\cite{LIGOScientific:2016lio,LIGOScientific:2020tif, LIGOScientific:2021sio}.
Unlike such RD models, \texttt{TEOBPM} includes information from the progenitors, allowing inference on parameters such as the initial masses and spins, and incorporates time-dependent models for the modes amplitudes tuned to NR, thus providing a bridge between inspiral-merger-ringdown (IMR) data and RD-only analyses.
This is similar in spirit to the strategy employed in Ref.~\cite{CalderonBustillo:2020rmh}.
Here, we improve on the latter by: 
i) using the complete time-domain likelihood, overcoming the circularity approximation; 
ii) using a time-domain model with an explicit parameterisation of all QNM-related quantities, allowing for a more direct physical interpretation of the constrains obtained; 
iii) incorporating the HMs content; 
iv) crucially including the uncertainty in the ringdown validity regime.

We report small positive Bayes factors (BF) in favour of the $(l,m)=(3,3)$ [$(l,m)=(2,1)$] mode in the events \GW{170729}, \GW{190521\_074359}, \GW{200129\_065458}, \GW{200224\_222234}\normalsize{} [\GW{190521\_074359}, \GW{191109\_010717}\normalsize{}], and study in details the robustness of the result by varying the starting time of the analysis.
We also introduce fractional shifts in the QNMs spectrum to study potential deviations from general relativity (GR) in the RD, discussing in detail tests of the final state conjecture under the assumption of the presence of two angular modes.
Our results support encouraging prospects for the positive observation of HMs in GW signals observed in the ongoing LVK observing run (O4)~\cite{KAGRA:2013rdx, LIGOScientific:2014pky, VIRGO:2014yos, PhysRevD.88.043007}.

% Summary of the paper
In Sec.~\ref{sec:rd_model}, we discuss the basics of multimodal RD waveforms from an observational perspective.
In Sec.~\ref{sec:TEOBPM_model}, we describe our implementation of the \texttt{TEOBPM} RD model.
In Sec.~\ref{sec:hms_detectability}, we introduce a semi-analytical procedure to predict the detectability of different HMs.
In Sec.~\ref{sec:gwtc3_analysis} we discuss the details of our Bayesian RD analysis and collect the parameter estimation (PE) results on GWTC-3, comparing them with the results obtained by the LVK Collaboration, both in the IMR and RD regimes.
We also describe our procedure to robustly search for HMs, including the systematic uncertainty in the ringdown starting time.
In Sec.~\ref{sec:TGR}, we introduce fractional shifts in the QNMs spectrum to study potential deviations from GR in the RD, discussing in detail tests of the final state conjecture under the assumption of the presence of two angular modes.
We conclude and discuss future prospects in Sec.~\ref{sec:conclusions}.

%==========================================================================
\section{Ringdown models}
\label{sec:rd_model}

% RD waveform
The linear gravitational perturbations of a Kerr BH are described by the Teukolsky equation~\cite{PhysRevLett.29.1114, 1973ApJ...185..635T}, with the main GW emission component captured by
\begin{equation}\label{linear_rd_wf}
\begin{split}
    h_{+} -i h_{\times} &= \frac{M}{d_L} \sum_{lm} h_{lm} Y_{lm}(\iota, \varphi),\\
    h_{lm} &\equiv A_{0,lm}\,e^{-\sigma_{lm}(t-t_0)+i\phi_{0,lm}},
\end{split}
\end{equation}
where $M \equiv m_1 + m_2$ is the total initial mass of the binary, $d_L$ is the luminosity distance, $Y_{lmn}$ are spin-weighted spherical harmonics (SWSH).
The \textit{plus} and \textit{cross} (gauge-invariant) polarizations of the GW waveform are denoted by $h_{+}$ and $h_{\times}$.
We define the QNMs complex frequencies as
\begin{equation}\label{eq:qnms_TEOBPM}
    \sigma_{lm}\equiv \alpha_{lm}+i\omega_{lm},
\end{equation}
where $\Re{\sigma_{lm}} = \alpha_{lm} \equiv 1/\tau_{lm}$ is the inverse of the \textit{damping time}, $\tau_{lm}$, while $\Im{\omega_n} = \omega_{lm}$ is the \textit{oscillation frequency}.
$A_{0,lm}$ and $\phi_{0,lm}$ are the \textit{constant} amplitudes and phases of the different modes.
Note that in Eq.~\eqref{linear_rd_wf} we are neglecting the \textit{overtone} expansion of each angular mode, since the contribution of overtones in the linear regime (where they can be confidently identified) is negligible compared to the longest-lived angular modes~\cite{London:2018nxs, Baibhav:2023clw}, on which we will focus.

We will refer to the mode $(l,m)=(2,2)$ as the \textit{fundamental mode}, and to those with $(l,m)\neq(2,2)$ as \textit{higher modes} (HMs).
The values of $\omega_{lmn}$ and $\tau_{lmn}$ are known semi-analytically and can be computed numerically given the mass and (dimensionless) angular momentum $\qty{M_f, a_f}$ of the final BH~\cite{Berti:2005ys, London:2018nxs, Cook:2014cta}.
On the other hand, $A_{0,lmn}$ depend on the specific process generating the perturbation, and are not known analytically for a binary merger of two BHs of comparable mass.
%
% Excitation of HMs
For example, for quasicircular equal-masses binaries, the dominant contribution in the RD is given by the fundamental mode $(2,2)$, with leading subdominant contributions coming from the modes $\qty{(3,3),(2,1),(3,2)}$, see Refs.~\cite{Kamaretsos:2011um, Kamaretsos:2012bs, London:2014cma,London:2018gaq, Bhagwat:2019bwv, Bhagwat:2019dtm, JimenezForteza:2020cve, Ota:2021ypb, Cheung:2022rbm}.
HMs are excited by asymmetries in the system~\cite{Berti:2007fi}, e.g. by increasing the mass ratio $q\equiv m_1/m_2$ ($m_1\geq m_2$) or the adimensional spins $\chi_{1,2}\equiv J_{1,2}/m_{1,2}^2$ of the two orbiting BHs.
In addition to the properties of the source, the actual content of the modes present in the RD also depends on the SWSHs, which further modulate different modes based on the geometry of the system.
In particular, the overall amplitude of the modes strongly depends on the inclination angle $\iota$, defined as the angle between the direction of the orbital angular momentum\footnote{Note that the orbital angular momentum direction coincides with the angular momentum of the final BH for non-precessing binaries.} and that of the observer.

The extraction of the QNM content from a GW signal, also known as ``black hole spectroscopy''~\cite{Detweiler:1980gk,Kokkotas:1999bd,Dreyer:2003bv,Berti:2009kk,Berti:2018vdi,Cardoso:2019rvt}, is performed using various sets of models similar to Eq.~\eqref{eq:qnms_TEOBPM}, which can be classified by the amount of information included (i.e. of assumptions).
The models with the fewest assumptions will be the most generic, capable of detecting even large deviations from GR predictions.
On the other hand, because they contain little information, such models provide less precise constraints on the deviation parameters and are the least sensitive to small GR deviations.
These classes of RD models range from pure superposition of damped sinusoids with unknown complex frequencies and amplitudes (i.e. with $\sigma_{lm}$ in Eq.~\eqref{eq:qnms_TEOBPM} an unknown parameter inferred from the data)~\cite{LIGOScientific:2016lio, Carullo:2019flw,LIGOScientific:2020tif, LIGOScientific:2021sio, Baibhav:2023clw}, 
to templates including only perturbative predictions on the QNM spectrum~\cite{Gossan:2011ha, Carullo:2019flw,Bhagwat:2019dtm,Bhagwat:2019bwv, Isi:2019aib, Isi:2021iql,Cabero:2019zyt,JimenezForteza:2020cve,Capano:2021etf, Cotesta:2022pci} but no information on the QNM amplitudes (i.e. $\sigma_{lm}$ constrained by GR predictions, but $A_{0,lmn}$ a free parameter in Eq.~\eqref{eq:qnms_TEOBPM}), 
to templates that incorporate numerical predictions of the quasinormal amplitudes (i.e. both $\sigma_{lm}$ and $A_{0,lmn}$ constrained by GR predictions).
The latter assume that the remnant black hole forms from a binary merger, and either incorporate the explicit amplitude dependence on binary parameters~\cite{Kamaretsos:2011um, Kamaretsos:2012bs,London:2014cma,London:2018gaq, Carullo:2018sfu, Hughes:2019zmt, Lim:2019xrb, JimenezForteza:2020cve, CalderonBustillo:2020rmh, Lim:2022veo, Zhu:2023fnf} or only partial information on the relative amplitudes excitation~\cite{Capano:2021etf}.
Furthermore, one can include numerical predictions for early-times non-linearities by constructing phenomenological models for the full postmerger signal by promoting $A_{0,lmn}$ to a time-dependent quantity, which is achieved by \texttt{TEOBPM}, but still without accounting for pre-merger data in the analysis.
The class of models with the largest amount of information (hence the most accurate, but less agnostic ones) are \texttt{pSEOB}-like templates~\cite{Brito:2018rfr,Ghosh:2021mrv,Maggio:2022hre,Silva:2022srr,Toubiana:2023cwr}, where even the pre-merger signal is included (currently under the hypothesis that pre-merger data are correctly described by GR), and deviations are allowed only in the QNM spectrum.
All these models should be regarded as complementary, and answering different questions.
For example, models that impose QNM spectral predictions might miss deviations induced by the non-Kerr nature.
Models that impose amplitudes predictions coming from binary inspirals may be biased if the orbital dynamics is not captured by quasicircular models (due to orbital eccentricity, precession, environmental effects etc.).
In particular, the \texttt{pSEOB} class of templates, within its current implementation, is expected to be the most sensitive to deviations that abruptly appear around the merger phase, e.g. triggered by high-curvature dynamical couplings, similar to dynamical scalarisation~\cite{Silva:2020omi}.
These classes of scenarios would predict an inspiral signal close to the GR prediction, but a merger-ringdown phase that is rather different.
Instead, searches based on ringdown-only templates that exclude pre-merger data, such as \texttt{TEOBPM} that is used in this study, will produce less biased results for scenarios where deviations are also present in the inspiral.
Clearly, the most sensitive and accurate test would be performed using coherent IMR models in specific alternative scenarios.
Recent progress has been made in constructing such models, both analytically~\cite{Julie:2017pkb,Julie:2017ucp, Jain:2022nxs, Julie:2022qux, Jain:2023fvt,Yunes:2011we,Blazquez-Salcedo:2016enn,Blazquez-Salcedo:2017txk,Blazquez-Salcedo:2020caw,Cardoso:2019mqo,McManus:2019ulj,Pierini:2021jxd,Pierini:2022eim,Srivastava:2021imr,Wagle:2021tam,Cano:2019ore,Adair:2020vso,Cano:2020cao,Cano:2021myl,Cano:2023tmv,Cano:2023jbk,Cano:2023qqm,Wagle:2023fwl,Li:2023ulk,Chung:2023zdq,Chung:2023wkd, Tattersall:2017erk,Franciolini:2018uyq} and numerically~\cite{East:2021bqk, Ripley:2022cdh, Corman:2022xqg, Evstafyeva:2022rve, Cayuso:2023aht}.
However, the large classes of possible deviations from GR makes this a daunting task, and naturally calls for generic models and parameterisations~\cite{Maselli:2019mjd,Carullo:2021dui,Yunes:2009ke,Cornish:2011ys,Chatziioannou:2012rf,Sampson:2013lpa,Sampson:2013wia,Loutrel:2022xok} capable of capturing deviations from several classes of modifications.

%==========================================================================
\section{The \texttt{TEOBPM} ringdown model}
\label{sec:TEOBPM_model}

% TEOBPM introduction
In this work we focus on a specific RD model, which constitutes the postmerger part of state-of-the-art EOB~\cite{Bohe:2016gbl,Cotesta:2018fcv} and phenomenological time-domain models~\cite{Estelles:2020osj,Estelles:2021gvs,Estelles:2020twz}.
The \textit{tidal effective one-body post-merger} (\texttt{TEOBPM}) model was first introduced in Ref.~\cite{Damour:2014yha}, as an NR-informed analytical representation of the postmerger waveform from BH coalescences.
The model has been further improved in~\cite{DelPozzo:2016kmd, Nagar:2018zoe, Nagar:2019wds, Nagar:2020pcj} as the postmerger part of the \texttt{TEOBResumS} waveforms family.
Here, we briefly review the model construction, and highlight its characteristics relevant to RD analyses.

% TEOBPM: waveform, nonlinearities, starting time
We work with the \textit{expansion coefficients} $h_{lm}$ in Eq.~\eqref{linear_rd_wf}, which are the output of NR simulations~\cite{Brown:2007jx}.
We want our model to be defined from the peak of the full IMR waveform, that we define as $t_0 \equiv t_{22}^{peak}$.
If the entire postmerger phase would be linear, we could directly apply Eq.~\eqref{linear_rd_wf} to describe the entire RD with an appropriate mode combination, and constant amplitudes~\cite{Baibhav:2023clw}.
Since the early times contain non-linearities, while Eq.~\eqref{linear_rd_wf} is obtained from \textit{linear} perturbation theory, additional contributions are required to extend the template to the peak of the waveform.
Thus, the strategy adopted in \texttt{TEOBPM} is to factorise the QNMs linear contribution from each mode $h_{lm}$, and define the \textit{QNM-rescaled ringdown waveform} $\bar{h}_{lm}$
\begin{equation}\label{rescaled_teob_wf}
    h_{lm} \equiv e^{-\sigma_{lm}(t-t_0)-i\phi_{0,lm}}\,\bar{h}_{lm} \, ,
\end{equation}
containing all the non-linear contributions.
To model non-linearities, we decompose the QNM-rescaled waveform $\bar{h}_{22}$ into a generic complex number, with time-dependent amplitude and phase,
\begin{equation}\label{rescaled_teob_wf_complex}
    \bar{h}_{22}(\tau) \equiv A_{\bar{h}_{22}}(\tau)e^{+i\phi_{\bar{h}_{22}}(\tau)} \, .
\end{equation}
The functions $A_{\bar{h}_{22}}$ and $\phi_{\bar{h}_{22}}$ are informed from NR simulations.
Here, we will not further describe the details of the procedure to produce these fits, and we refer the interested reader to Refs.~\cite{Nagar:2019wds, Nagar:2020pcj}.

% HMs
The above procedure is applied to each mode.
Nevertheless, the time at which different modes peak will not necessarily be the same.
That is, in general, $t_{lm}^{\text{peak}}$ will not coincide with the merger time, defined as the peak of the fundamental mode $t_{0}=t_{22}^{\text{peak}}$.
The \textit{time delay} $\Delta t_{lm}$ is then defined as the difference between the peak of the fundamental mode $(2,2)$ and the peak of each mode
\begin{equation}
    \Delta t_{lm} \equiv t_{lm}^{\text{peak}}-t_{22}^{\text{peak}} \equiv t_{lm}^{\text{peak}}-t_{0} \, .
\end{equation}
Consequently, for each HM included in the model, one additional parameter $\Delta t_{lm}$ needs to be fit on NR simulations.
From Eq.~\eqref{linear_rd_wf}, the expansion coefficients for each mode are then
\begin{multline}\label{TEOBPM_multimode_template}
    h_{lm}(t+\Delta t_{lm}) = e^{-\sigma_{0,lm}t-i\phi_{0,lm}}\,\bar{h}_{lm}\\
    =e^{-\sigma_{0,lm}t-i\phi_{0,lm}}\,A_{\bar{h}_{lm}}(t+\Delta t_{lm})\: e^{+i\phi_{\bar{h}_{lm}}(t+\Delta t_{lm})} \, .
\end{multline}
Further details on our implementation of \texttt{TEOBPM} can be found in App.~\ref{sec:appendix_a}.

% TEOBPM characteristics
\subsection{Characteristics of \texttt{TEOBPM}}\label{subsec:TEOBPM_characteristics}
Having discussed how \texttt{TEOBPM} is constructed and how some of its internal degrees of freedom are fixed from NR, we list the remaining free parameters of the model.
% TEOBPM parameters
Being $\Theta_{lm}$ the parameter space on which \texttt{TEOBPM} is defined, $\Theta_{lm}$ is parametrized by $\vb*{\theta}=\qty{\vb*{\theta}_{I},\vb*{\theta}_{E}} \in \Theta_{lm}$, where
\begin{equation}\label{TEOBPM_params}
\begin{split}
    &\textit{intrinsic}:\quad \vb*{\theta}_{I} \;\equiv\qty{m_1, m_2, \chi_1, \chi_2} \, ,\\
    &\textit{extrinsic}:\quad \vb*{\theta}_{E}\equiv \qty{\alpha, \delta, \psi, \iota, d_L, t_0, \phi_{0,22}, \phi_{0,lm}} \, .
\end{split}
\end{equation}
Thus, \texttt{TEOBPM} is a ringdown model with 11 parameters when only the fundamental mode is used, and one additional parameter $\phi_{0,lm}$ for each HM included. Note, however, that some of the extrinsic parameters are typically fixed from IMR results in ringdown-only analyses either because of degeneracies or to fix the analysis segment, so that the actual number of free parameters is smaller.
Similarly, the orbital phase $\varphi$ entering the spherical harmonics is not included due to the complete degeneracy with single-modes phases.
This differs from usual RD models, which are typically parametrized in terms of remnant properties, such as the final mass and dimensionless spin $\qty{M_f, a_f}$ or the QNMs complex frequencies $\qty{\omega_{lm}, \tau_{lm}}$.
This parametrization allows \texttt{TEOBPM} to describe ringdown characteristics in terms of the progenitors properties.
These choices make the model particularly sensitive to GR deviations, as discussed below, at the cost of being less agnostic about the nature of the RD process, a fact that is particularly relevant for tests of general relativity (TGR).\\
%
%
% TEOBPM advantages and limitations
Below, we highlight the advantages and the limits of \texttt{TEOBPM}.\\
Advantages:
\begin{itemize}
    \item \textit{Starting time}\quad The model is defined from the peak of the fundamental mode $(2,2)$, which can be estimated from IMR analyses, either modelled~\cite{LIGOScientific:2019fpa} or unmodelled~\cite{Carullo:2018gah}.
    Consequently, the uncertainty in determining the starting time of our RD analysis depends only on the width of the IMR peaktime posterior distribution.
    Instead, more agnostic analyses based on superposition of damped sinusoids that do not include near-peak contributions, additionally require to identify a reference time at which the pure-QNM description of the data starts to be valid~\cite{Carullo:2018sfu,Bhagwat:2017tkm}.
    Since the system is evolving and thus the GW frequency is changing with time (as the remnant mass and spin, see e.g.~\cite{Baibhav:2023clw}), a pure-QNM description will never be exact.
    Hence, for pure damped-sinusoids, an adequate reference time will be chosen in such a way that the systematic uncertainties due to non-stationary QNM contributions are much smaller than the statistical uncertainties due to the finite SNR.
    At present detectors sensitivity, depending on the accuracy of the model, this happens in the range $[10,15] \, M$~\cite{Carullo:2018sfu,LIGOScientific:2020tif,LIGOScientific:2021sio}.

    \item \textit{Ampitudes}\quad Since the initial amplitudes $A_{0,lm}$ of the different modes are set by a fit to NR, the model includes the information on the excitation of the various modes as a function of the progenitor parameters (similar to e.g. the \texttt{Kerr_{HMs}} model~\cite{London:2014cma,London:2018gaq} used in Refs.~\cite{LIGOScientific:2020ufj,LIGOScientific:2020iuh,LIGOScientific:2020tif,LIGOScientific:2021sio}, also labeled ``\texttt{MMRDNP}'' in \texttt{pyRing}).
    
    \item \textit{Non-linearities}\quad As the model includes the entire postmerger part of the signal, it uses more data with high SNR with respect to linear RD models, providing more sensitive results (compared to e.g. the ``\texttt{MMRDNP}'' model~\cite{London:2014cma,London:2018gaq}). 
    This makes \texttt{TEOBPM} particularly suitable for characterising HMs.
\end{itemize}
Limitations:
\begin{itemize}
    \item \textit{Precession} and \textit{eccentricity}\quad The model is informed on NR simulations of quasicircular binary BHs with spins aligned or anti-aligned with the orbital angular momentum.
    As a result, it is only valid for events in which both precession and eccentricity are negligible.
    
    \item \textit{Mode-mixing}\quad When the gravitational radiation is decomposed into SWSHs instead of spin-weighted \textit{spheroidal} harmonics, modes with the same $m$ and different $l$ are affected by mode mixing~\cite{Buonanno:2006ui, Kelly:2012nd, Berti:2014fga}.
    Among the most relevant modes, the $(3,2)$ mixes with the fundamental mode, and this effect is not currently included in \texttt{TEOBPM}.
    Although we expected the contribution of the mode $(3,2)$ to be negligible at current sensitivity, see App.~\ref{sec:appendix_d}, this issue should be addressed in the future.
    This contribution has recently been included in the v5 version of the SEOB model~\cite{Pompili:2023tna,vandeMeent:2023ols,Khalil:2023kep,Ramos-Buades:2023ehm,Mihaylov:2023bkc}, applied to spectroscopic analyses in the context of LISA in Ref.~\cite{Toubiana:2023cwr}.
    
    \item \textit{Accuracy}\quad The accuracy of the model has been extensively studied on NR simulations, with some of the results reported in App.~\ref{sec:appendix_b}.
    We observe single-mode mismatches of $O(10^{-3})$ for the fundamental mode $(2,2)$, and $O(10^{-2})$ for the $\qty{(3,3),(2,1)}$ modes, in agreement with the literature~\cite{Nagar:2020pcj}.
    Others subdominant modes $\qty{(3,2),(4,4),(4,3),(4,2)}$ show a worse performance, but their overall contribution to the waveform can be safely ignored at current sensitivity. Hence, we decided not to include these modes in the analysis and leave a systematic study of their inclusion to future work.
    Note that since the $(2,2)$ is by far the dominant mode in the systems under consideration, the mismatch of the total waveform (summing all the modes) receives only small contributions from subdominant modes, so the total mismatch is typically close to that of the fundamental mode.
    
    \item \textit{$\Delta t_{lm}$}\quad As the HMs start at different times compared to the fundamental, a ringdown-only implementation of the model is systematically missing the contribution of HMs during the time interval $\Delta t_{lm}$, i.e. until the mode has started.
    Note that in the corresponding IMR version, this part of the signal comes from the inspiral dynamics and is therefore present.
    This results in an intrinsic limitation in the current implementation of \texttt{TEOBPM} in \texttt{pyRing}.
    Although we have found this effect to be negligible at current sensitivities based on the analysis of simulated signals, it may have an impact for future studies at higher sensitivities. It could be addressed, for example, by constructing time-dependent amplitude fits starting from a common peak time for all HMs.
    
    \item \textit{$\phi_{0,lm}$}\quad The presence of an additional parameter for the initial phase of each HM impacts the BF by increasing the prior volume, systematically penalising the hypotheses with higher number of modes (Ockham’s razor).
    This problem could be solved by fixing the initial phases on NR in future refinements of the model.
    This implies that our current implementation leads to more conservative results when computing BFs for the presence of HMs.
\end{itemize}

%==========================================================================
\section{Higher modes detectability}
\label{sec:hms_detectability}

\begin{figure*}
    \hspace*{-0.5cm}\includegraphics[scale=0.84]{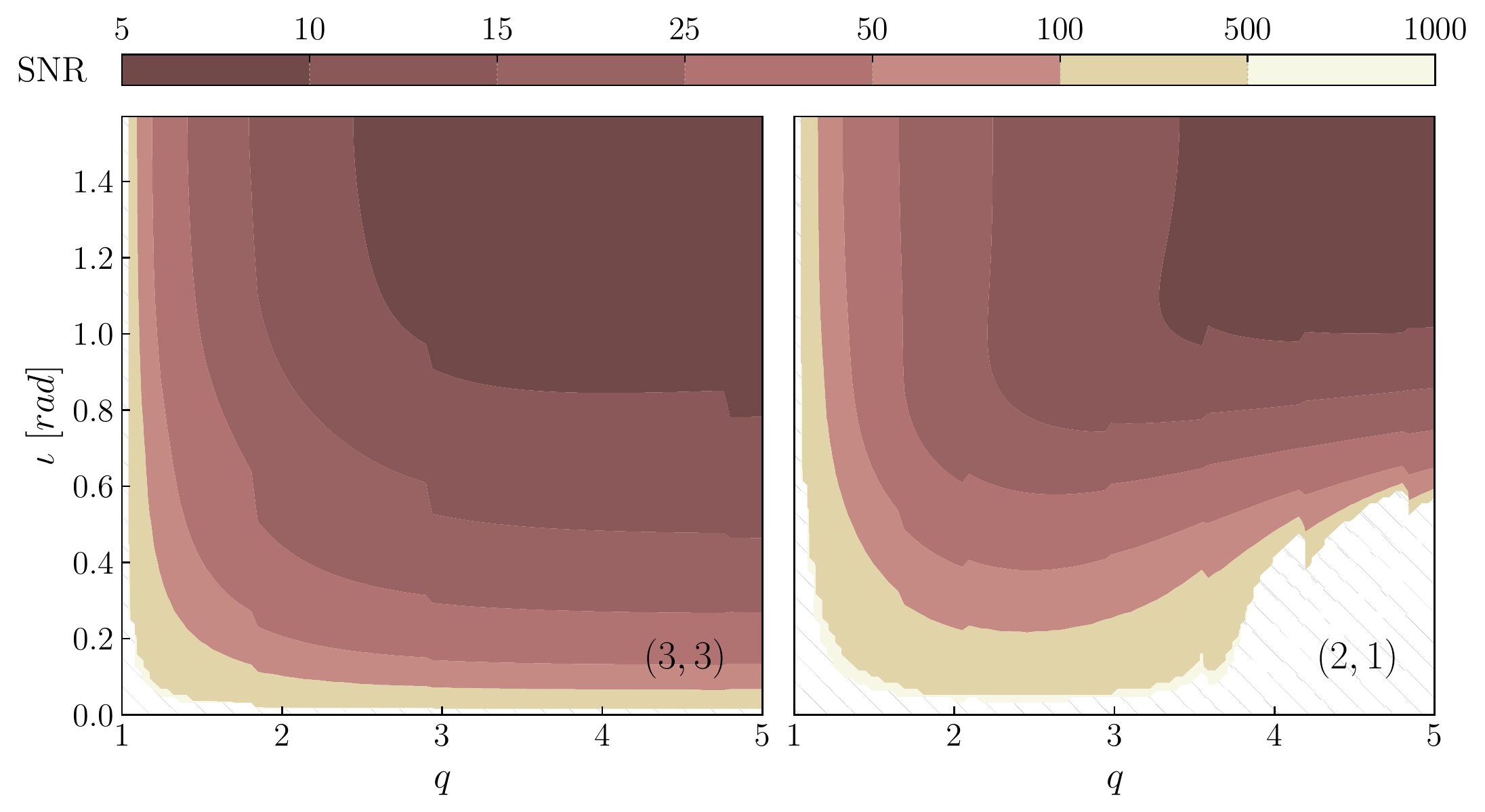}
    \caption{\footnotesize Colormap of the optimal signal-to-noise ratio needed to measure a logarithmic Bayes factor of $\ln{\mathcal{B}_{lm,22}}=3$, in favour of the mode $(3,3)$ (left) and $(2,1)$ (right) over the fundamental mode. The results are for two nonspinning initial black holes $\chi_1=\chi_2=0$, and are expressed as function of the mass ratio $q$ and inclination $\iota$.
    The dashed regions correspond to $\text{SNR}>1000$.}
    \label{fig:snr_33-21}
\end{figure*}
%

% Introduction to the problem
We now turn to the investigation of HMs observability in the RD of current GW events.
To answer this question, we develop a semi-analytical procedure to predict the detectability of the different HMs over the fundamental mode $(2,2)$.
These techniques have been first introduced in the context of TGRs to study the detection of GR deviations in the high SNR limit~\cite{Cornish:2011ys, Vallisneri:2012qq, DelPozzo:2014cla}. \newline

% Fitting factor and SNR
In general, we define the time-domain scalar product weighted with the inverse \textit{autocovariance matrix} as
\begin{equation}\label{eq:scalar_product_td}
    \bra{\vb{a}}\ket{\vb{b}} \equiv \vb{a}^T \mathbb{C}^{-1}_{\vb{n}\vb{n}} \bar{\vb{b}},
\end{equation}
where $\vb{a}$ and $\vb{b}$ are two time series, and $\mathbb{C}_{\vb{n}\vb{n}}$ is the autocovariance matrix of the detector noise.
Specifically, we use an inverse Fourier Transform of the simulated power spectral density (PSD) from the \textit{advanced LIGO design} configuration~\cite{noiseadv} to estimate the autocovariance matrix. 
Note that the FF only depends on the profile of the PSD, as Eq.~\eqref{eq:fitting_factor} in invariant under any rescaling of the autocovariance matrix.
The \textit{fitting factor} (FF) is then defined as
\begin{equation}\label{eq:fitting_factor}
    \text{FF} \equiv \max_{\vb*{\theta}\in\Theta_{lm}} \qty{\frac{\braket{\vb{h}_{lm}}{\vb{h}_{22}}^2}{\braket{\vb{h}_{lm}}{\vb{h}_{lm}}\braket{\vb{h}_{22}}{\vb{h}_{22}}}} \, ,
\end{equation}
where $\vb*{\theta}$ are the parameters of the model, and $\Theta_{lm}$ is the parameter space over which the model including HMs is defined.
As proven in App.~\ref{sec:appendix_c}, the FF is closely related to the optimal signal-to-noise ratio ($\text{SNR}_{\text{opt}}$), see App.~\ref{sec:appendix_c} for the definition of the latter quantity.

% Detectability equation
Suppose we have measured some data $\vb{d}$, and want to understand whether a given HM  present in the signal can be distinguished from the fundamental mode.
This is a \textit{model selection} problem between the following hypotheses:
\begin{itemize}
    \item $\mathcal{H}_{22}$: in the signal only the fundamental mode $(2,2)$ is present;
    \item $\mathcal{H}_{lm}$: besides the fundamental mode $(2,2)$, the signal contains a given HM $(l,m)$.
\end{itemize}
As shown in App.~\ref{sec:appendix_d}, we can derive an equation that relates the Bayes factor $\mathcal{B}_{lm,22}$ between the two hypotheses, the SNR, and the FF
\begin{equation}\label{bayes_factor_snr_fitting_factor}
    \ln{\mathcal{B}_{lm,22}} = \frac{1}{2}\qty(1-\text{FF}^2)\, \text{SNR}_{\text{opt}}^2,
\end{equation}
which is valid for large SNRs.
Eq.~\eqref{bayes_factor_snr_fitting_factor} can be used to analytically predict the detectability of different HMs as a function of the parameters of the waveform model.
In fact, given a set of parameters for the system, we can use the waveforms $\vb{h}_{22}$ and $\vb{h}_{lm}$ to calculate the FF through Eq.~\eqref{eq:fitting_factor}, and then Eq.~\eqref{bayes_factor_snr_fitting_factor} provides the value of $\ln{\mathcal{B}_{lm,22}}$ as a function of $\text{SNR}_{\text{opt}}$.
This quantifies how much the hypothesis $\mathcal{H}_{lm}$ is favoured over $\mathcal{H}_{22}$, depending on the power of the signal $\vb{h}_{lm}$ over noise.
In other words, Eq.~\eqref{bayes_factor_snr_fitting_factor} answers the following question:
\begin{itemize}
    \item[] \textit{what is the signal-to-noise ratio needed to reach some threshold value of the Bayes factor in favour of higher modes?}
\end{itemize}
In practice, we set the threshold value for a detection to be $\ln{\mathcal{B}_{lm,22}}=3$, and invert Eq.~\eqref{bayes_factor_snr_fitting_factor} to find $\text{SNR}_{\text{opt}}$ as a function of the $\text{FF}$.
The FF varies across the binary parameter space, because the excitation of the HMs depends on the progenitor parameters.
Hence, we can study the SNR needed for the detection as a function of the parameters of the model.
Fig. \ref{fig:snr_33-21} shows, for nonspinning initial black holes $\chi_1=\chi_2=0$, the $\text{SNR}_{\text{opt}}$ required to measure the modes $(3,3)$ and $(2,1)$, using \texttt{TEOBPM} as waveform model.
The SNR is given as a function of the mass ratio $q\in [1,5]$ and inclination $\iota\in [0,\pi/2]$, the latter interval being sufficient for spin-aligned systems due to equatorial symmetry.

For equal mass binaries $q\sim 1$, such as those typically observed in current events~\cite{LIGOScientific:2021djp}, we do not expect to observe HMs in the RD.
Indeed, with a threshold $\ln{\mathcal{B}_{lm,22}}=3$, we would need an event with $\text{SNR}_{\text{opt}} \gtrsim 50$ to detect the mode $(3,3)$, with current events having $\text{SNR}_{\text{opt}} \lesssim 15$ in the RD.
However, HMs are excited by introducing asymmetries in the system, and increasing the mass ratio dramatically reduces the threshold SNR.
For example, with $q\in \qty[2,3]$ we could confidently measure the mode $(3,3)$ for loud events at current sensitivities $\text{SNR}_{\text{opt}}\sim \qty[10,15]$, for inclinations $\iota > \pi/4$.
Note that the mode $(2,1)$ generally has SNRs larger than the mode $(3,3)$, especially for low mass ratios $q \lesssim 2$.
This implies that the $(3,3)$ mode is the most easily observable HM in the RD.
Similar plots for the modes $\qty{(3,2),(4,4)}$ can be found in App.~\ref{sec:appendix_d}, showing that the SNR required to observe their contribution is larger compared with the modes $\qty{(3,3),(2,1)}$.
For instance, we need $\text{SNR}_{\text{opt}} \gtrsim 25$ to observe the mode $(4,4)$ at mass ratios $q \lesssim 2$ and inclinations $\iota > \pi/4$.
Note that the above numbers depend strongly on the choice of the threshold BF which, if decreased, suggests a potential for detection already at low mass ratios.
Going beyond our conservative threshold would require simulation studies in real interferometric noisy data, where the threshold would be chosen to keep the probability of noise-induced detection below a certain value.
Importantly, these simulation studies would also need to account for modelling systematics due to the imperfect representation of the NR solutions, and for the uncertain determination of the starting time from IMR signals.

To ensure numerical stability when inverting Eq.~\eqref{bayes_factor_snr_fitting_factor} to evaluate the SNR, we set a numerical cutoff around $\text{FF}=1$ at the level of $10^{-5}$.
Therefore, regions where $\text{FF}=1$ at that level of precision are displayed in white in Fig.~\ref{fig:snr_33-21}.
The sharp edges in the level curves are induced by the profile of the PSD used in the scalar product of Eq.~\eqref{eq:fitting_factor}.
Finally, we stress that the detectability results just discussed are derived under the approximation of high SNR, $\vb{h} \gg \vb{n}$, and should therefore be used as a rough estimate and complemented by parameter estimation studies for cases of specific interest.

%==========================================================================
\section{GWTC-3 analysis}
\label{sec:gwtc3_analysis}

% Introduction
We now describe the RD analysis with \texttt{TEOBPM} on real data from the latest catalog of GWs, GWTC-3.
The study was performed with \texttt{pyRing}~\cite{pyRing}, a \texttt{python} package for time domain RD analyses of BH coalescences, based on the formalism described in~\cite{DelPozzo:2016kmd,Carullo:2019flw, Isi:2019aib} and Sec. VII.A.1 of~\cite{LIGOScientific:2020tif}.
For a detailed and pedagogical introduction to time-domain analyses, see Ref.~\cite{Isi:2021iql}.
To compute Bayesian evidences and probability distributions, we use the \texttt{raynest}~\cite{raynest} sampler, running each analysis with 4 parallel chains, 5000 live points and 5000 maximum number of steps in the Markov Chain Monte Carlo (MCMC).

\subsection{Input parameters and events selection}
\label{sec:evts_selection}
% Preliminary analysis
In a first study, we select the subset of events that present an informative RD with the \texttt{TEOBPM} model.
To select such events, we compute $\ln{\mathcal{B}_{s,n}}$ between the hypotheses:
\begin{itemize}
    \item $\mathcal{H}_{s}$:\quad the data contain a GW signal $\vb{h}$ modeled with only the fundamental mode, $\vb{d}=\vb{h_{22}}+\vb{n}$;
    \item $\mathcal{H}_{n}$:\quad the data only consist of noise $\vb{n}$, $\vb{d}=\vb{n}$;
\end{itemize}
and then select the events with $\ln{\mathcal{B}_{s,n}}>3$.
Following standard procedures in time-domain RD analyses, we fix some of the free parameters of \texttt{TEOBPM} from existing PE analyses in the full IMR regime, to fix the analysis segment.
Such procedure prevents the template from latching to pre-peak data portions, and and has been verified to introduce negligible biases at the current sensitivity~\cite{LIGOScientific:2021sio}.
In particular, the peak time $t_0$ and the sky position $\qty{\alpha, \delta}$ are taken from the LVK TGR papers~\cite{LIGOScientific:2019fpa, LIGOScientific:2020tif, LIGOScientific:2021sio}.
For repeatability, the input values used in this work are reported in App.~\ref{sec:appendix_e}.
The remaining free parameters are listed in Tab. \ref{prior_table}, together with the prior ranges used, uniform in all the quoted variables.
The prior ranges on masses and spins are taken from~\cite{Schmidt:2020yuu}.

% Preliminary analysis results
We apply this procedure to the 48 events included in the GWTC-3 catalog based on their \textit{false alarm rate}, as described in~\cite{LIGOScientific:2019fpa, LIGOScientific:2020tif, LIGOScientific:2021sio}.
Applying the above procedure, we select a subset of 18 events that are informative in the RD, which are those that will be used in the rest of the work, listed in Tab.~\ref{tab:pe_TEOBPM}.
\begin{center}
    \begin{tabular}{ ccccccc }
        \hline
        $m_{1,2}$ & $\chi_{1,2}$ & $\phi_{0,22}$ & $\ln{d_L}$ & $\cos{\iota}$ & $\varphi$ & $\psi$ \\
        \small{$[10,200]$} & \small{$[-0.8,0.95]$} & \small{$[0,2\pi]$} & \small{$[\ln{10}, 10]$} & \small{$[-1,1]$} & \small{$[0,2\pi]$} & \small{$[0,\pi]$} \\
        \hline
    \end{tabular}
    \captionof{table}{\footnotesize Free parameters used in the analysis, with their prior range. The masses are expressed in solar masses \SI{}{M_{\odot}} and the luminosity distance in mega-parsecs \SI{}{Mpc}.}
    \label{prior_table}
\end{center}
%

% Parameter estimation and LVK comparison
\subsection{Parameter estimation and LVK comparison}

\begin{figure*}[htp]
    \begin{subfigure}{\textwidth}
    \includegraphics[width=\textwidth]{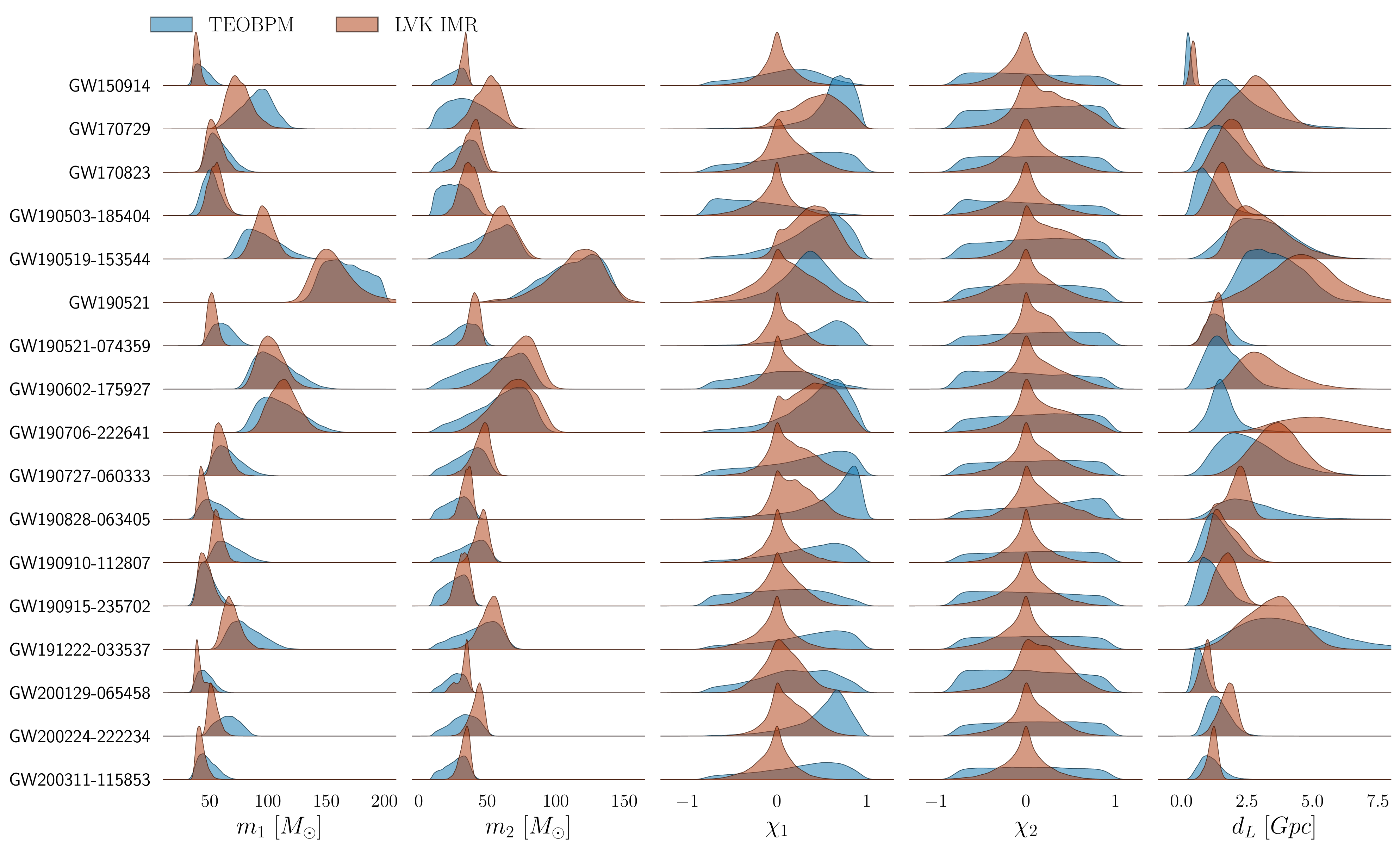}
    \end{subfigure}
    \begin{subfigure}{\textwidth}
    \includegraphics[width=\textwidth]{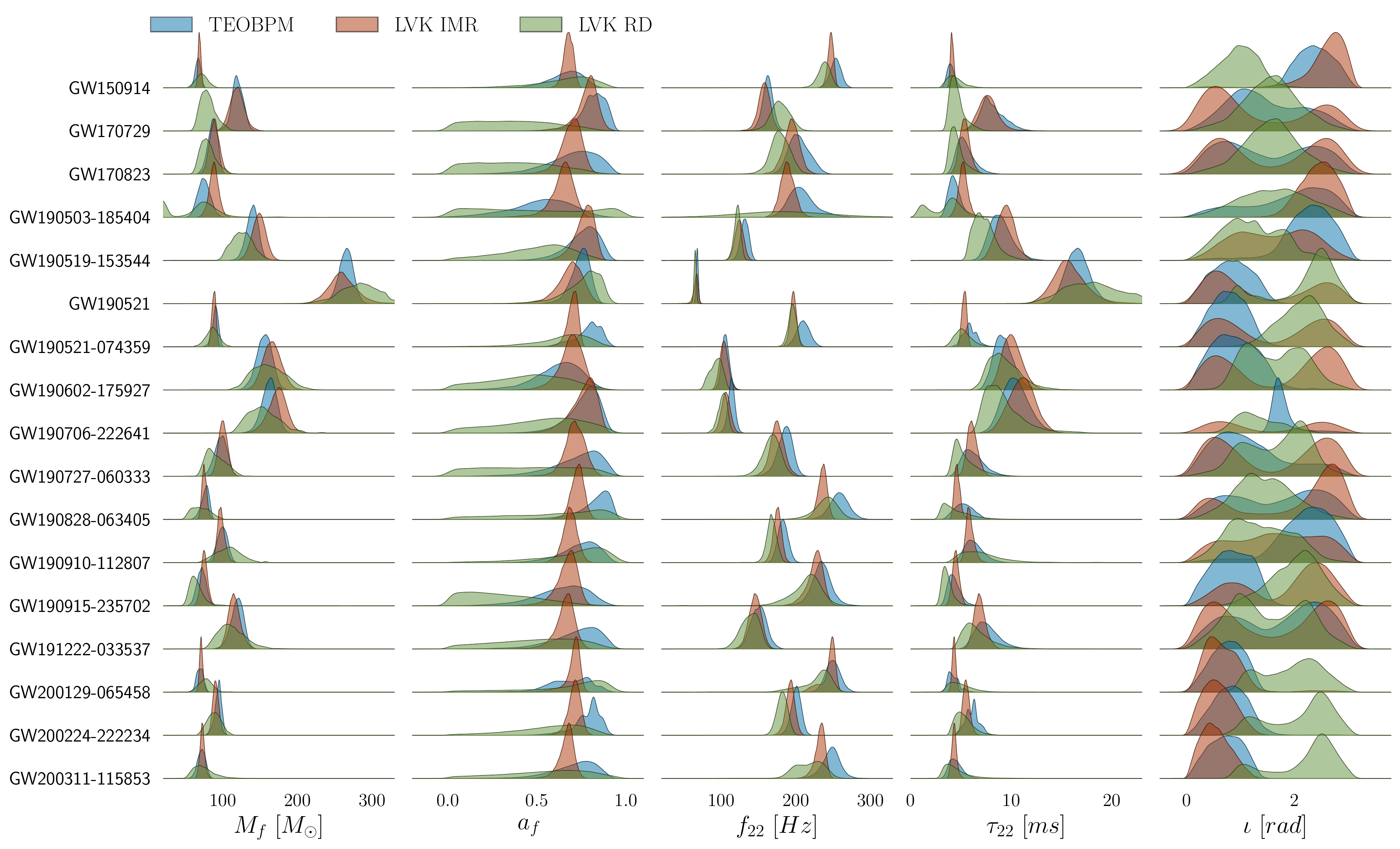}
    \end{subfigure}
    \caption{\footnotesize
    Ridgeline plot with the posterior distributions of the following parameters: primary mass $m_1$, secondary mass $m_2$, primary adimensional spin $\chi_1$, secondary adimensional spin $\chi_2$, luminosity distance $d_L$, final mass $M_f$, final spin $a_f$, frequency and damping time of the fundamental mode, $f_{22}$ and $\tau_{22}$, inclination $\iota$.
    Each row correspond to one of the events analysed.
    The \textit{blue} distributions have been obtained with \texttt{TEOBPM}, the \textit{red} distributions are taken from IMR analyses by the LVK Collaboration, the \textit{green} distributions are from RD analyses by the LVK Collaboration.
    In agreement with the Bayes Factors of the multimodal analysis, for the events \GW{170729}\footnotesize{}, \GW{190521\_074359}\footnotesize{}, \GW{200129\_065458}\footnotesize{}, \GW{200224\_222234}\footnotesize{} we use the PE from the analysis with the modes $\qty{(2,2), (3,3)}$, while for the other events with only the fundamental mode.}
    \label{fig:violin_LVK-TEOBPM_b}
\end{figure*}

We perform PE with the \texttt{TEOBPM} on the selected events and compare the results with those from the LVK Collaboration, both in the IMR and RD regime.
As motivated in the previous section, for \texttt{TEOBPM} we consider only the modes $(l,m) = [(2,2), (3,3), (2,1)]$.
Fig.~\ref{fig:violin_LVK-TEOBPM_b} shows the marginalized posterior distributions for a comprehensive list of parameters, and Tab. \ref{tab:pe_TEOBPM} summarises the median values of the posteriors obtained with \texttt{TEOBPM} for selected parameters.
The LVK IMR samples are combined from different waveform models, including precessing templates.
%
% Ridgeline TEOBPM-LVK
The LVK RD samples are taken from the \texttt{pyRing} analysis using the \texttt{Kerr_{221}} template, as presented in the LVK GWTC-3 TGR catalog~\cite{LIGOScientific:2021sio}, starting at $t_0=t_{22}^{\text{peak}}$ like \texttt{TEOBPM}.
The \texttt{Kerr_{221}} model assumes a Kerr remnant BH and the fundamental mode with its first overtone.
Overtones close to the peak~\cite{Giesler:2019uxc, Isi:2019aib} do not correspond to physical vibrational frequencies of the underlying spacetime~\cite{Baibhav:2023clw}, but their use allows the entire post-peak signal to be fitted with sufficient accuracy for current (low-SNR) GW detections.
Hence, the overtone was exploited as an effective term to push the analysis at earlier times, capturing more power present in the signal.
This allows a direct comparison with the \texttt{TEOBPM} results, as both analyses include the same amount of data.
Note, however, that the two models adopt two different parameterisations.
While \texttt{TEOBPM} depends on the progenitor parameters, the \texttt{Kerr_{221}} depend on remnant parameters, such as $\qty{M_f, a_f, A_{lmn}, \phi_{lmn}}$.
For \texttt{TEOBPM} and LVK IMR, the remnant parameters $\qty{M_{f}, a_{f}, f_{22}, \tau_{22}}$ are obtained with fits on NR data~\cite{Jimenez-Forteza:2016oae, Berti:2005ys}.
The posteriors shown are generated using the kernel density estimation (KDE) method in \texttt{scipy}.

Results are generally consistent across the different models.
We note that, in general, \texttt{TEOBPM} obtains distributions that are sharper than LVK RD and broader than LVK IMR.
This behaviour is expected, as the IMR analysis uses more data and recovers higher SNRs.
Nonetheless, in most cases \texttt{TEOBPM}'s posteriors are comparable to those of LVK IMR, and in certain cases sharper in some parameters.
There are several possible reasons for this behaviour.
A first possibility is related to the fact that IMR models also include pre-peak data, so that noise features could contribute to produce different posteriors.
This could be either because noisy features present in the inspiral data are not included in the RD analysis, or because IMR analyses are less sensitive to noise fluctuations due to the larger SNR.
In addition, the LVK IMR samples are obtained with precessing models, and precession introduces a strong correlation between the parameters, broadening the marginalised posteriors.
For this reason, since we assume aligned spins in \texttt{TEOBPM}, the spins comparison with LVK IMR should not be expected to yield identical results.
Finally, neglecting the correlation between sky position and luminosity distance, as done in the RD analyses, will also artificially sharpen the posteriors.
We leave a more detailed investigation of these features to future work.
Instead, \texttt{TEOBPM} contains more information in the model compared to LVK RD, and will therefore be able to extract more SNR if the signal is well described by GR.
Furthermore, sampling on the initial masses and spins $\qty{m_{1,2}, \chi_{1,2}}$ instead of remnant parameters $\qty{M_f, a_f, A_{lmn}, \phi_{lmn}}$ further reduces correlations, thus sharpening the \texttt{TEOBPM}'s posteriors compared to LVK RD.

\begin{table}
    \setlength{\tabcolsep}{5pt}
    \centering
    \scalebox{0.82}{
    \hspace*{-0.1cm}\begin{tabular}{ lllllll }
        \hline
        \hline
        event & \normalsize{$\,m_1$} & \normalsize{$\,m_2$} & $\chi_1$ & $\iota$ & $d_L$ \\
        & \footnotesize $[\SI{}{M_{\odot}}]$ & \footnotesize $[\SI{}{M_{\odot}}]$ & & \footnotesize $[\text{rad}]$ & \footnotesize $[\SI{}{Gpc}]$ \\
        \hline
        \GW{150914}         & $ 42.9_{ -6.5}^{ +9.2}$ & $ 27.0_{-11.1}^{ +7.3}$ & $0.3_{-0.2}^{+0.3}$ & $2.4_{-0.4}^{+0.4}$ & $0.3_{-0.1}^{+0.1}$ \\[1mm]
        \GW{170729}         & $ 84.3_{-18.0}^{+25.2}$ & $ 43.7_{-23.6}^{+16.7}$ & $0.7_{-0.4}^{+0.2}$ & $1.7_{-1.2}^{+1.0}$ & $2.7_{-1.4}^{+2.0}$ \\[1mm]
        \GW{170823}         & $ 56.8_{ -9.4}^{+14.3}$ & $ 33.7_{-15.4}^{+10.6}$ & $0.4_{-0.4}^{+0.4}$ & $1.4_{-1.0}^{+1.2}$ & $1.5_{-0.7}^{+1.0}$ \\[3mm]

        \GW{190503\_185404} & $ 50.9_{ -8.9}^{+10.4}$ & $ 26.1_{-12.3}^{+12.7}$ & $0.4_{-0.3}^{+0.3}$ & $2.1_{-1.3}^{+0.7}$ & $1.0_{-0.5}^{+0.8}$ \\[1mm]
        \GW{190519\_153544} & $ 92.6_{-15.9}^{+23.4}$ & $ 54.8_{-26.7}^{+16.2}$ & $0.5_{-0.4}^{+0.3}$ & $2.3_{-0.6}^{+0.5}$ & $3.0_{-1.4}^{+1.6}$ \\[1mm]
        \GW{190521}         & $165.7_{-21.0}^{+25.8}$ & $114.9_{-30.4}^{+21.9}$ & $0.4_{-0.3}^{+0.3}$ & $0.8_{-0.5}^{+0.5}$ & $3.4_{-1.2}^{+1.5}$ \\[1mm]
        \GW{190521\_074359} & $ 59.7_{-10.1}^{+14.4}$ & $ 35.4_{-16.0}^{+10.3}$ & $0.6_{-0.4}^{+0.3}$ & $0.8_{-0.5}^{+0.4}$ & $1.4_{-0.6}^{+0.7}$ \\[1mm]
        \GW{190602\_175927} & $103.4_{-16.9}^{+25.1}$ & $ 58.8_{-31.1}^{+21.1}$ & $0.3_{-0.3}^{+0.4}$ & $0.9_{-0.5}^{+0.6}$ & $1.5_{-0.7}^{+1.0}$ \\[1mm]
        \GW{190706\_222641} & $108.1_{-19.1}^{+27.2}$ & $ 63.4_{-30.8}^{+19.1}$ & $0.5_{-0.4}^{+0.3}$ & $1.7_{-0.2}^{+0.4}$ & $1.5_{-0.6}^{+0.8}$ \\[1mm]
        \GW{190727\_060333} & $ 64.2_{-11.5}^{+17.3}$ & $ 37.7_{-17.5}^{+12.0}$ & $0.5_{-0.4}^{+0.3}$ & $1.1_{-0.7}^{+1.3}$ & $2.5_{-1.2}^{+1.8}$ \\[1mm]
        \GW{190828\_063405} & $ 52.3_{-10.8}^{+14.2}$ & $ 28.3_{-12.8}^{+10.1}$ & $0.7_{-0.5}^{+0.2}$ & $1.8_{-1.3}^{+0.9}$ & $2.4_{-1.1}^{+1.7}$ \\[1mm]
        \GW{190910\_112807} & $ 65.3_{-11.6}^{+17.6}$ & $ 38.8_{-18.1}^{+11.8}$ & $0.5_{-0.4}^{+0.3}$ & $2.3_{-0.7}^{+0.5}$ & $1.4_{-0.6}^{+0.8}$ \\[1mm]
        \GW{190915\_235702} & $ 46.9_{ -7.6}^{+10.9}$ & $ 27.6_{-12.4}^{ +9.3}$ & $0.4_{-0.3}^{+0.4}$ & $0.8_{-0.5}^{+0.5}$ & $1.1_{-0.5}^{+0.8}$ \\[3mm]

        \GW{191109\_010717} & $128.1_{-28.2}^{+25.2}$ & $ 50.1_{-32.6}^{+31.5}$ & $0.8_{-0.2}^{+0.1}$ & $1.8_{-0.1}^{+0.3}$ & $0.8_{-0.4}^{+0.8}$ \\[1mm]
        \GW{191222\_033537} & $ 79.3_{-13.9}^{+20.7}$ & $ 46.1_{-22.7}^{+14.8}$ & $0.5_{-0.4}^{+0.3}$ & $1.9_{-1.3}^{+0.8}$ & $3.8_{-1.9}^{+2.6}$ \\[1mm]
        \GW{200129\_065458} & $ 44.6_{ -6.9}^{ +9.7}$ & $ 27.8_{-11.9}^{ +7.6}$ & $0.3_{-0.3}^{+0.4}$ & $0.8_{-0.5}^{+0.4}$ & $0.7_{-0.3}^{+0.3}$ \\[1mm]
        \GW{200224\_222234} & $ 61.4_{-10.6}^{+14.6}$ & $ 36.5_{-16.6}^{+10.8}$ & $0.5_{-0.4}^{+0.3}$ & $0.9_{-0.5}^{+0.4}$ & $1.4_{-0.6}^{+0.7}$ \\[1mm]
        \GW{200311\_115853} & $ 46.7_{ -8.0}^{+11.9}$ & $ 28.3_{-12.5}^{ +8.2}$ & $0.4_{-0.3}^{+0.3}$ & $0.7_{-0.4}^{+0.4}$ & $1.1_{-0.5}^{+0.6}$ \\[1mm] 
        \hline
        \hline
    \end{tabular}}
    \captionsetup{width=0.5\textwidth, format=plain, justification=raggedright, singlelinecheck=false}
    \captionof{table}{\footnotesize Median and $90\%$ symmetric credible intervals of some relevant parameters: primary mass $m_1$, secondary mass $m_2$, dimensionless primary spin $\chi_1$, inclination $\iota$, luminosity distance $d_L$.}
    \label{tab:pe_TEOBPM}
\end{table}

The initial spins are the parameters which vary the most compared to LVK IMR, with \texttt{TEOBPM} systematically recovering higher values of $\chi_{1}$.
However, we note that the IMR pointy distributions in $\chi_{1,2}$ are mostly dominated by the prior, which is different from ours.
In fact, we use a flat prior in $\chi_{1,2}\in \qty[-0.8,0.95]$, while IMR analyses typically have three free components for each spin, and set a prior uniform in spin magnitude on the sphere. See~\cite{LIGOScientific:2021djp}.
Indeed, we highlight how the agreement with our results improves dramatically in the few events where the IMR posteriors on $\chi_{1,2}$ differ from the prior, such as \GW{170729}, \GW{190519\_153544}\normalsize{} and \GW{190706\_222641}\normalsize{}.
To understand whether these high values of the primary spin could be induced by detector noise, we simulated several mock signals at different SNRs, and found no biases in the recovery of the injected value.
We also investigated how our flat prior in the component masses affects the prior in the effective spin through correlations with the (non-flat) mass ratio, finding that such an effect is not sufficient to justify the results.

\begin{figure*}
    \includegraphics[width=\textwidth, scale=1.1]{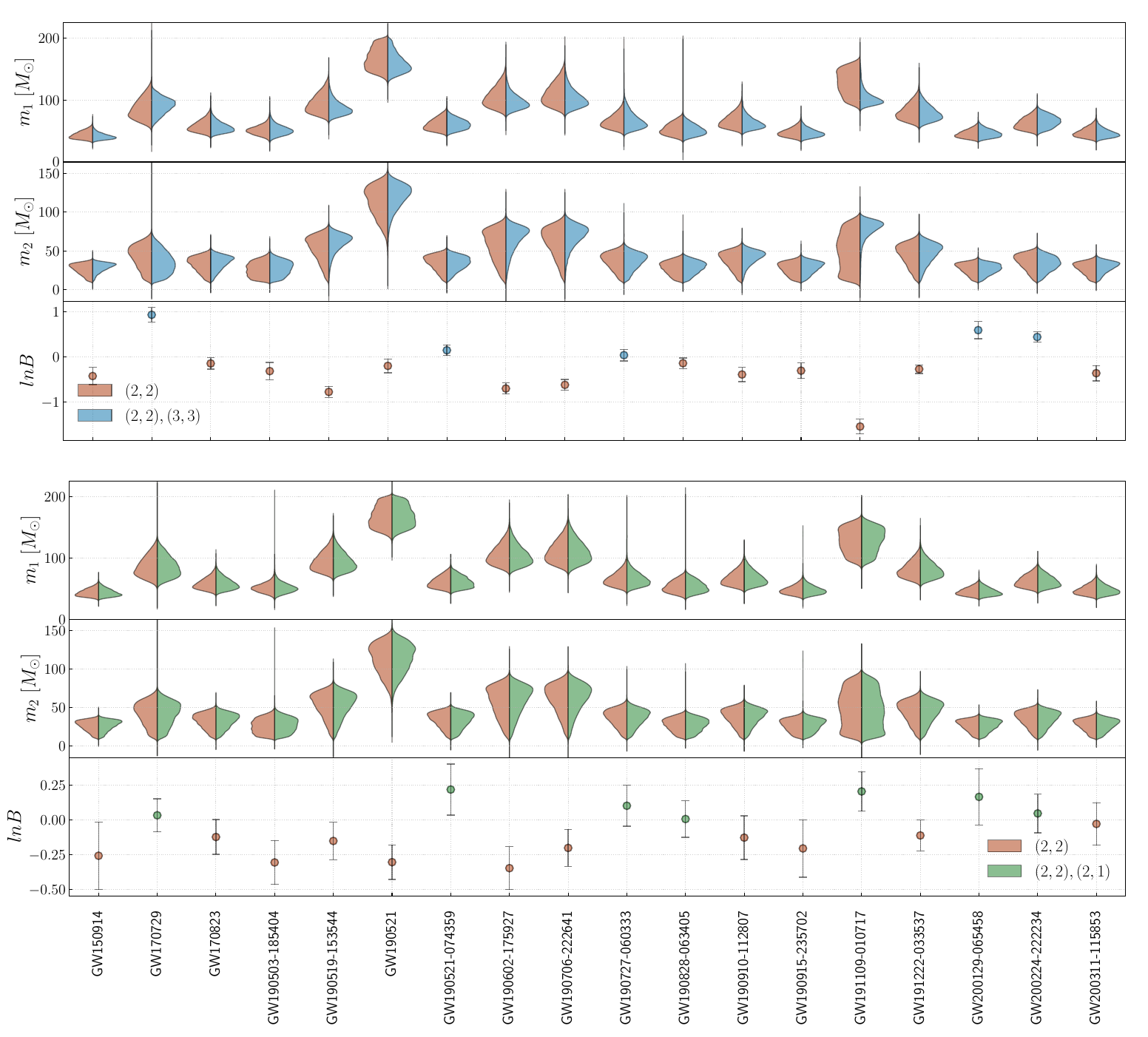}
    \captionsetup{format=plain, justification=raggedright, singlelinecheck=false}
    \caption{\footnotesize Posteriors distributions of the initial masses $m_{1,2}$ for the different events on the x-axis.
    The \textit{red} posteriors are from the analysis with only the fundamental mode $(2,2)$, the \textit{blue} ones using the fundamental mode and the mode $(3,3)$, the \textit{green} ones using the fundamental mode and the mode $(2,1)$.
    For the two subplots, the bottom row shows the logarithmic Bayes factor $\ln{\mathcal{B}_{lm,22}}$ between the HM considered and only the fundamental mode, colored according to which hypothesis is favoured.}
    \label{fig:HMs_search}
\end{figure*}
%

% Multimodal analysis
\subsection{Multimodal analysis}\label{sec:multimodal_analysis}

In the previous section we compared ringdown-only estimates to full IMR analyses, when assuming the $(l,m) = [(2,2), (3,3), (2,1)]$ modes.
These results can be interpreted as a time-domain version of the IMR consistency test~\cite{Ghosh:2016qgn,Ghosh:2017gfp,Breschi:2019wki,Carson:2020cqb,Islam:2021pbd,LIGOScientific:2021sio,Bhat:2022amc}, similar to what was done in Ref.~\cite{Isi:2020tac}, although here we do not perform a corresponding inspiral analysis and only compare to full IMR results.
For an application to astrophysical source properties using similar techniques, see Ref.~\cite{Miller:2023ncs}.

Now, we turn our attention to the detection of modes beyond the fundamental, with the goal of extracting multiple QNM modes from the system, allowing for tests of GR predictions with a more immediate interpretation.
To this end, we conduct a multimodal search to investigate the presence of HMs in the RD of the selected events.
Based on the predictions on the detectability in Sec. \ref{sec:hms_detectability}, we only use the $\qty{(3,3),(2,1)}$ HMs and consider the following hypotheses:
\begin{itemize}
    \item $\mathcal{H}_{22}$:\quad\quad\quad $\qty{(2,2)}$;
    \item $\mathcal{H}_{22,33}$:\quad\quad $\qty{(2,2),(3,3)}$;
    \item $\mathcal{H}_{22,21}$:\quad\quad $\qty{(2,2),(2,1)}$;
    \item $\mathcal{H}_{22,33,21}$:\quad $\qty{(2,2),(3,3),(2,1)}$;
\end{itemize}
where each logical proposition represents the assumption of having a gravitational wave signal $\vb{h}$ in the data $\vb{d}$ with that specific set of modes.
We then perform model selection between these hypotheses, expressing the results through the logarithmic BF $\ln{\mathcal{B}_{lm,22}}$, where the model $lm$ corresponds to one of the above hypotheses including at least one HM.

% Results
Results are summarized in Tab. \ref{tab:multimodal_analysis}, where we report the BFs between the templates with and without HMs.
To gauge the impact of the priors, we also show the values of the network optimal signal-to-noise ratio $\text{SNR}^{\text{net}}_{\text{opt}}$, the information $H$~\cite{Skilling2006NestedSF} and the maximum value of the logarithmic likelihood $\text{lnL}_{\text{max}}$.
None of the events shows strong preference for HMs.
Nonetheless, the events \GW{170729}, \GW{190521\_074359}, \GW{200129\_065458}, \GW{200224\_222234}\normalsize{} have a low but positive BF.
To appreciate the consequences of this fact, in Fig. \ref{fig:HMs_search} we study how the PE on the initial masses $m_{1,2}$ varies when HMs are included.
In the case of the mode $(2,1)$, all the logarithmic BFs are very close to zero, and there are no appreciable differences in the posteriors with the additional mode.
Instead, in the case of the mode $(3,3)$, we can clearly see how the distributions change for the events which show a positive BF for this HM.
In particular, for these events the posteriors move towards higher mass ratios when mode $(3,3)$ is included, a behaviour that is qualitatively in agreement with the physical excitation of this mode.

\begin{figure*}
    \includegraphics[scale=1.05]{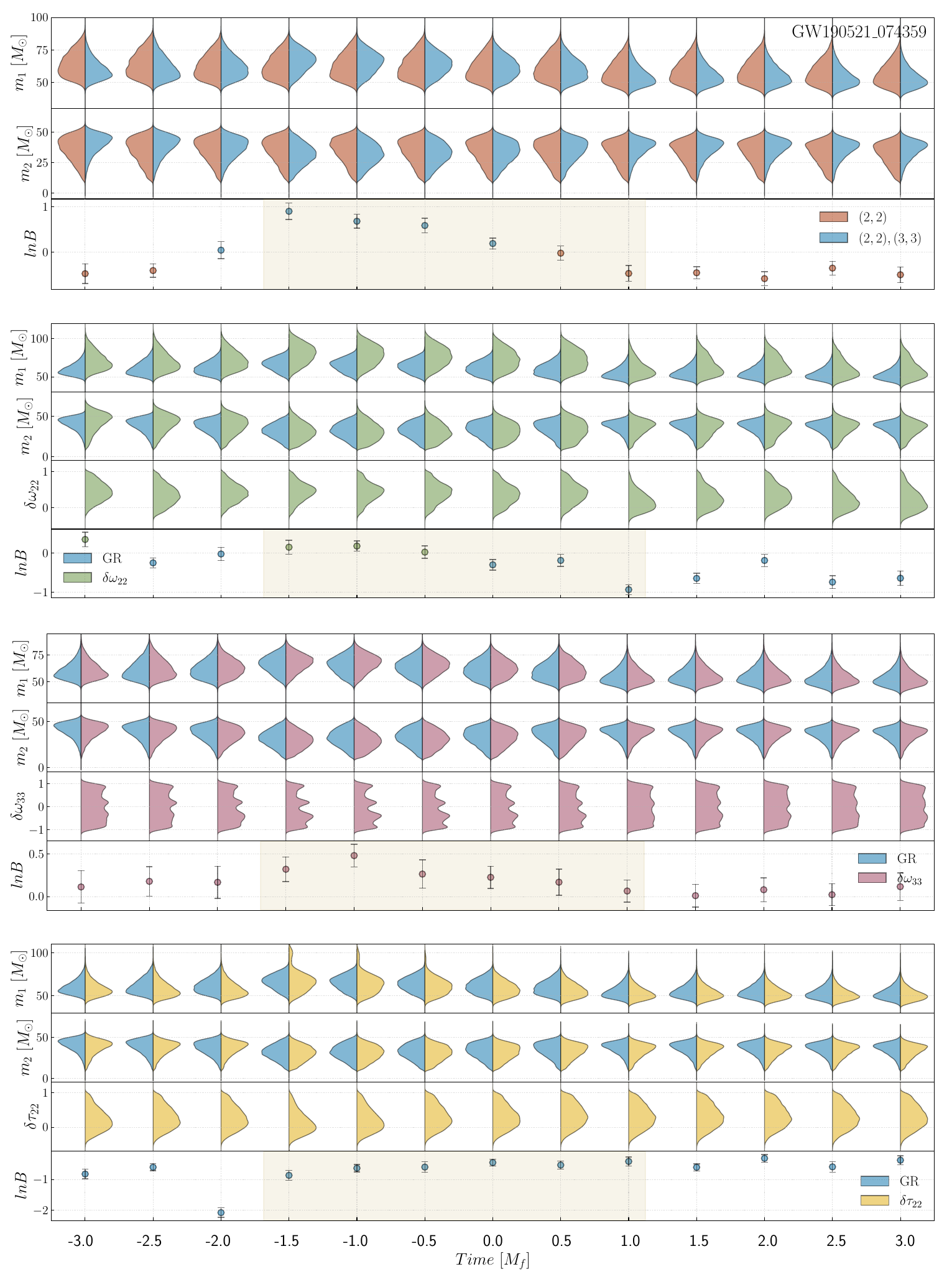}
    \caption{\footnotesize PE and model selection on the event \GW{190521\_074359}\footnotesize, as a function of the starting time.
    For each subplot, each column correspond to a separate analysis with a different starting time.
    The \textit{red} posterior distributions are obtained with only the fundamental mode $(2,2)$, the \textit{blue} ones with the modes $\qty{(2,2), (3,3)}$.
    For each subplot, the first two rows display the initial masses $m_{1,2}$, and the bottom row the logarithmic Bayes factor between the competing hypotheses, with a color corresponding to the favoured hypothesis.
    The shaded region in the bottom row outlines the $95\%$ CI of the corresponding IMR peaktime distribution.
    The subplot in the top correspond to the GR analysis, while the underline three include deviations from GR in the form of fractional deviations in the QNMs.
    Specifically, the \textit{green} distributions are obtained with the modes $\qty{(2,2), (3,3)}$ and deviations only on the frequency of the fundamental mode $\delta\omega_{22}$, the \textit{pink} ones with the deviations only in the frequency of the mode $(3,3)$, and the \textit{yellow} ones with deviations only in the damping time of the fundamental mode $\delta\tau_{22}$.}
    \label{fig:GW190521_074359}
\end{figure*}

% HMs at different starting times
However, these results are not yet sufficient to conclude a $(3,3)$ detection, as the search was only performed at one fixed starting time, given by the median value of the IMR peaktime distribution for each event.
This procedure is reasonable to keep the computational cost of the search under control, but needs to be extended if aiming to claim a mode detection.
In fact, since the peaktime distribution is an additional parameter, its uncertainty needs to be appropriately included (i.e. marginalised over) in the inference to obtain unbiased results.
It is well known that RD analyses are highly sensitive to the choice of the starting time, and a naive treatment of this issue can lead to non-robust or even highly biased results~\cite{Carullo:2019flw, Cheung:2022rbm, Baibhav:2023clw,Cheung:2023vki}.
This problem typically manifests itself in the following forms:
\begin{itemize}
    \item Moving to earlier times with respect to the model's prescription implies the use of data that are outside the regime of applicability of the model.
    In practice, for a \textit{linear} RD model, this normally results in the use of data containing non-linearities from the postmerger in an attempt to obtain more data with higher SNR.
    For \texttt{TEOBPM}, this means including in the data some premerger part of the signal before the peak of the waveform.
    In both cases, the first part of the data is affected by un-modelled low frequencies at high SNR, pushing the final mass to larger values.
    We also note that in the case of \texttt{TEOBPM}, there is also a bias when starting at later times after the peak.
    This is because we are assuming given values of the amplitudes and phases evaluated at the peak time.
    Applying them at a different reference time will result in a bias.
    This situation does not occur for RD models consisting of a superposition of QNMs with free amplitudes, where starting at later times only implies a loss in SNR.
    
    \item With a fixed starting time, it is not possible to evaluate the impact of short-transient detectors-noise realizations.
    Given the low-SNR contained in the post-merger only signal, gauging the impact of such effects becomes significantly more important.
    By repeating the analysis at multiple times, it is possible to trace certain features present in the BF or posterior to noise artifacts.
\end{itemize}
\texttt{TEOBPM} is defined to start at the peak of the IMR waveform.
Thus, we should probe the entire region where the IMR peaktime distribution has support.
This procedure is still dependent on the specific IMR model used, and results may vary for different IMR waveforms, but for events where the systematic uncertainties of the IMR modelling are under control, this typically has a small impact.
In the top subplot of Fig.s~\ref{fig:GW190521_074359},\ref{fig:GW170729},\ref{fig:GW200129_065458},\ref{fig:GW200224_222234} we show how the BFs and the $m_{1,2}$ posteriors change as the starting time is varied across the $95\%$ CI of the IMR peaktime distribution, for the subset of events that prefer the $(3,3)$ mode.
All the four events display similar features:
\begin{itemize}
    \item one or more peaks in the BF within the $95\%$ CI of the IMR peaktime distribution;
    \item a change in the posterior distributions favouring higher mass ratios in correspondence with a BF peak.
\end{itemize}
We interpret this behaviour as resulting from the impact of the mode $(3,3)$ in these events, which allows to improve the PE in the region of applicability of our model.
Note that this effect is visible despite the low evidence for the HM, suggesting that the posterior distributions are more sensitive than BFs in targeting features in the data, although the latter constitute a more conservative and robust metric~\cite{Laghi:2020rgl}.

For the events \GW{200129\_065458}\normalsize{} and \GW{200224\_222234}\normalsize{}, the BF continues to increase at late times outside the support of the peaktime distribution.
This effect is difficult to interpret due to the expected increase of modelling systematics when starting at times much later than the peak, and should be investigated in the future.
Likely cause scould to be the impact of systematics introduced by different IMR models in reconstructing the peak time, or precessional degrees of freedom not included in our analysis~\cite{LIGOScientific:2021djp, Hannam:2021pit, Payne:2022spz, Islam:2023zzj, Puecher:2023rxw, Macas:2023wiw}.

Finally, we discuss our results for the event \GW{190521}\normalsize{}, in which several works have investigated the presence of the mode $(3,3)$ in its RD~\cite{LIGOScientific:2020ufj,LIGOScientific:2020iuh, LIGOScientific:2020tif, Capano:2021etf, Capano:2022zqm, Siegel:2023lxl}.
In the upper subplot of Fig. \ref{fig:GW190521}, we observe a peak in the BF as function of the starting time, which is however shifted from the median value of the peaktime distribution and lies outside the $95\%$ CI.
At our resolution of $1M_f$, the BF reaches its maximum at $t_0 = -6 M_f = \SI{-7.6}{ms}$, with $M_f \simeq \SI{258.8}{M_{\odot}}$.
Our input parameters, including the peaktime distribution, were obtained with the \texttt{NRSur7dq4} IMR waveform.
Our results are in qualitative agreement with~\cite{Capano:2021etf, Siegel:2023lxl}, since also in those analysis significant evidence for HMs could only be obtained well outside the validity regime of the employed models.
Apart from the systematics induced by the start time, precession or eccentricity are known to bear a large impact on this event~\cite{Siegel:2023lxl,Romero-Shaw:2020thy,Gayathri:2020coq,Gamba:2021gap,CalderonBustillo:2020xms}, so further studies will be needed to properly characterise HMs detections in this type of systems.

Assessing the significance for the weak evidence we obtained for HMs, as quantified by BFs, would in principle require simulation studies~\cite{Capano:2022zqm}.
In such a study, a large number of NR signals containing only the (2,2) mode or the full HM content would be added on top of both gaussian and real detector noise (the latter would especially be required for low-significance detections), and recovered using our model.
Then, setting a threshold minimising the number of false positives detection of HMs when simulating a signal containing only the (2,2) mode, would imply a ``detection threshold'' for the BF.
Since here we do not claim significant evidence for HMs, we do not perform such a study, which would bear a large computational cost.
We note however, that gathering experience from a selected number of these simulations, systematics due to modelling and start time determination typically bear a much stronger impact on false positive detections.

%==========================================================================
% GW190521_074359

%==========================================================================
\section{Tests of general relativity}
\label{sec:TGR}

The final-state conjecture states that astrophysical BHs are uniquely described by the Kerr metric~\cite{Loutrel:2020wbw,Carullo:2019flw,Penrose1982SOMEUP,Ginzburg,Zeldovich,Zeldovich2,Israel,Carter,Hawking1972,Robinson,Bunting, Mazur}.
In our Bayesian framework, this is an \textit{assumption} nested within our waveform model that can to be verified against observations.
We can use the RD to test this hypothesis (commonly referred to as ``no-hair'' conjecture), by performing model selection between GR and non-GR models.
Additional degrees of freedom (``hairs'') of the final BH system are modelled by deviation parameters with respect to the GR prediction, typically in the QNM
spectrum.
Different parameterizations of the deviations can be used, depending on the degrees of assumptions built into the model.
For example, for RD models composed of pure QNM superpositions, at least two different modes need to be resolved to remove the degeneracy between $\qty{f_{22}, \tau_{22}}$ and $\qty{M_f, a_f}$, as discussed e.g. in Refs.\cite{Dreyer:2003bv, Li:2011cg, Gossan:2011ha}.
In modern analyses, this is typically done using overtones to phenomenologically model the early postmerger, see~\cite{Isi:2019aib,LIGOScientific:2020tif,LIGOScientific:2021sio}.
Instead, if the RD model used has amplitudes fixed from numerical relativity simulations (e.g.~\cite{Kamaretsos:2011um,Kamaretsos:2012bs,London:2014cma,London:2018gaq,Hughes:2019zmt, Lim:2019xrb, JimenezForteza:2020cve,Lim:2022veo,Cheung:2023vki}), the numerical calibration will impose a constraint on the parameter space which naturally decouples $\qty{f_{22}, \tau_{22}}$ and $\qty{M_f, a_f}$.
In this way the no-hair conjecture can be tested with only a single mode, at the price of introducing additional GR inputs in the model, i.e. at the price of a less agnostic search.
This is the case of \texttt{TEOBPM}, and similar considerations hold for the LVK analysis with \texttt{pSEOBNR}~\cite{LIGOScientific:2021sio}.
Nonetheless, even though \texttt{TEOBPM} allow us to perform such tests with only the fundamental mode, we expect the test to improve when multiple modes are observed.
We stress that the above considerations are only qualitative, and the actual situation on real data strongly depends on the SNR and characteristics of the system.

We introduce \textit{fractional deviations} in the oscillation frequencies of the modes as
\begin{equation}\label{eq:effective_frequencies}
    \omega_{lm} = \omega_{lm}^{\text{\scriptsize{GR}}} (1 + \delta\omega_{lm}),
\end{equation}
where $\omega_{lm}^{\text{\scriptsize{GR}}}$ is the value predicted by GR.
The same procedure is applied to the damping time $\tau_{lm}$.
Note that for \texttt{TEOBPM} the deviations are introduced directly in Eq.~\eqref{TEOBPM_multimode_template}.

For example, if only the fundamental mode is considered, the deviations can in general affect:
\begin{itemize}
    \item only the frequency $\delta\omega_{22}$;
    \item only the damping time $\delta\tau_{22}$;
    \item both the frequency and damping time.
\end{itemize}
In the presence of multiple modes, the number of possibilities quickly increases, as all combinations between the deviations in frequencies and damping times for the different selected modes have to be considered.
The systematic study of all these combinations is already challenging with two modes, in order to collect and compare the different results, in addition to the large computational cost.

\subsection{TGR with two modes}\label{sec:TGR_2modes}

We first test the set of events with positive (albeit weak) evidence for the mode $(3,3)$ and consider the following hypotheses:
\begin{equation}
    \mathcal{H}_{\delta\omega_{22}},\quad \mathcal{H}_{\delta\omega_{33}},\quad \mathcal{H}_{\delta\tau_{22}},\quad \mathcal{H}_{\qty{\delta\omega_{22}, \delta\tau_{22}, \delta\omega_{33}}}.
\end{equation}
The baseline GR model consists of the modes $\qty{(2,2), (3,3)}$, and each hypothesis is constructed from this basic model with the deviation parameters contained in the subscript of each hypothesis.
We neglect the deviations $\delta\tau_{33}$, since damping times (and so their deviations) are in general more difficult to be measured than the frequencies.
As in Sec. \ref{sec:multimodal_analysis}, we repeat the analysis by varying the starting time across the $95\%$ CI of the IMR peaktime distribution.
This amounts to ``marginalise'' a-posteriori over the start time uncertainty.
The results are shown in Fig.s~\ref{fig:GW190521_074359},\ref{fig:GW170729},\ref{fig:GW200129_065458},\ref{fig:GW200224_222234}.
The marginalized distributions for the deviations are consistent with GR, i.e. they have support on zero up to the $90\%$ CI, except in the cases discussed below.
The BFs also show no strong preference for the non-GR hypotheses, with the largest BF in favour of GR deviations being $\mathcal{B} \simeq 3$ in the support of the peak time distribution.
This happens for the subset of events with lower SNR and higher masses, which merge at frequencies where the noise variance rises sharply, so such low-significance results are to be expected when analysing a large number of events at many different starting times, just based on the statistical properties of the noise.
Outside of the peak time support, the BFs tend to grow, sometimes to significantly large values.
However, as discussed previously, the model assumes that the start time coincides with the waveform peak, hence spurious GR deviations are expected when this assumption is violated significantly.
A procedure to combine results at multiple starting times into a single estimate would help to immediately assess the impact of this effect.
Such a procedure would re-weight the obtained BFs according to the value of the peak time probability distribution~\footnote{We thank Roberto Cotesta for first suggesting this algorithm at the ``Black-Hole Ringdown Workshop'' held at CCA, Flatiron Institute, New York (2022).}, combining the various BF$_{t_{st}=t_i}$, into a unique BF$_w$, where $i$ runs on the time samples overlapping with the $t_p$ support.
A similar algorithm has been implemented in Ref.~\cite{Finch:2022ynt}.
A crucial difference is that the latter analysis assumed a pre-merger model, allowing such marginalisation to be easily performed.
Within time-domain methods, the analysis segment is selected by excising all data before $t_{st}$, so analyses performed at different $t_{st}$ will include different portions of data.
This implies that combining different BFs would produce a BF$_w$ which is not evaluated on a fixed amount of data, hence cannot be straightforwardly interpreted as a proper BF, in the sense of standard probability rules.
More refined procedures will be required to produce such a combined BF$_w$.
However, an upper bound on BF$_w$ can always be constructed by taking  $\rm BF_{max} \coloneqq max \,\, BF_{t_i}$, which is the approach we follow here.
In the following, we briefly discuss our results, to illustrate the main features observed in the posteriors and BFs.

\begin{itemize}
    \item \GW{170729}\normalsize{}\quad The GR value is mostly included in the support of the deviation posteriors, although for a few cases at the boundaries of the peak time posterior, the GR value lies only in the tails of the distribution, especially when considering deviations on the decay time.
    This is consistent with $t_{st}$ induced systematics at times far away from the probable peak, as discussed above.
    Additionally, since the event has a relatively low $\text{SNR}^{\text{net}}_{\text{opt}}\sim 5.7$, noise-induced deviations are expected.
    This is reflected in bimodalities in the posteriors, both in the deviation parameters and $\{m_{1,2}, \chi_1, d_L\}$, which vary inconsistently over time.
    The bimodalities manifest in a peak consistent with GR, and a secondary peak which appears at the lower prior bound (equal to -1) of the deviation parameters $\delta\omega_{22}$ and $\delta\tau_{22}$.
    This secondary peak correlates with high mass ratios and low distances.
    Such behaviour indicates that the signal can be fitted equally well by the higher mode (significantly enhanced for high mass ratio), with the fundamental mode suppressed (the lower bounds of the deviation parameters correspond to a very fast decaying signal, with very low frequency outside of the detector sensitive band), and is another manifestation of the low SNR.
    Concerning the upper bound, we prefer to keep the deviations bounds fixed to the (already significantly wide) interval $[-1,1]$, since railing is expected to hold even when the bounds are increased, as for all signals where little information is present, as verified e.g. in Ref.~\cite{Ghosh:2021mrv}.

    \item \GW{190521\_074359}\normalsize{}\quad The posteriors on $\delta\omega_{22}$ in the time interval $[-1.5,0.5]M_f$ are slightly shifted towards larger values, but the GR prediction always lies well within the $90\%$ CI.
    This effect is compensated by the increase in the mass $m_1$, which is positively correlated with $\delta\omega_{22}$.
    As expected, the BF also increases slightly in the same time interval.
    $\delta\tau_{22}$ always favours GR, and $\delta\omega_{33}$ is poorly resolved\footnote{Even if the results are not shown here, we checked that this behaviour hold when the three deviations are inferred \textit{together}, that is $\mathcal{H}_{\qty{\delta\omega_{22}, \delta\tau_{22}, \delta\omega_{33}}}$. The same is applies to the other events.}.

    \item \GW{200129\_065458}\normalsize{}\quad The deviation posteriors are always centered on the GR value, with the BFs always favouring GR.
    Such an improved agreement is expected given the large value of $\text{SNR}^{\text{net}}_{\text{opt}}\sim 12.5$, which also allows for deviations on $\delta\omega_{33}$ to be slightly resolved.
    The deviation observed in Ref.~\cite{Maggio:2022hre} are due to waveform systematics when including the pre-merger signal, which do not appear here.
    This is likely because of the reduced impact of precession on the post-merger signal.

    \item \GW{200224\_222234}\normalsize{}\quad 
    The deviations $\delta\omega_{22}$ and $\delta\tau_{22}$ are centered on GR, with BFs favouring GR accordingly.
    Also for this event, the deviations $\delta\omega_{33}$ are partially resolved, with a peak emerging, although the support of the posterior does not exclude any region of the prior, hence results are essentially uninformative.
\end{itemize}
Note how, in all the events, the deviations $\delta\omega_{22}$ try to capture the premerger part of the signal at times far away from $t_p$, with large variations in the posteriors accompanied by high non-GR BFs, indicating a weak  consistency of the model with the data.
This interesting, and expected, behaviour serves to remind how sensitive the results of RD analyses are to the underlying assumptions, and how important is the proper handling of $t_{st}$.
For completeness, Fig. \ref{fig:GW190521} reports the deviations in \GW{190521}\normalsize{}, that are well measured and centered on the GR value for $\qty{\delta\omega_{22}, \delta\omega_{33}}$.
This is particularly appreciable in correspondence with the BF peak that lies outside the IMR peaktime distribution, as discussed in Sec. \ref{sec:multimodal_analysis}.
However, we observe that the posteriors in $\delta\tau_{22}$ are systematically shifted toward higher values with respect to the GR expectation.
This behaviour is accompanied by a positive BF for non-GR in the entire support of the peaktime distribution.
Further studies on the GR deviations with only the fundamental mode will be necessary to draw conclusions.

Finally, Fig. \ref{fig:GW191109_010717}\normalsize{} shows the results for the event \GW{191109\_010717}\normalsize{}, for which previous LVK analyses have identified the presence of non-GR deviations induced by non-stationarities in the noise~\cite{LIGOScientific:2021sio}.
We reproduce this result, and label this event as the showcase of a non-robust analysis, as manifest by the highly varying posteriors.

\subsection{TGR with the fundamental mode}
\label{sec:TGR_fundamental}

We summarise in Fig. \ref{fig:TGR_fundamental} the results on the events with no evidence of HMs, for which we test the hypotheses $\mathcal{H}_{\delta\omega_{22}},$ and $\mathcal{H}_{\delta\tau_{22}}$.
For simplicity, we perform the analysis at only one starting time, given by the median value of the IMR peaktime distribution.
For the first gravitational signal \GW{150914}\normalsize{}, among the loudest ringdown observed to date, we instead provide the complete analysis at different starting times in Fig. \ref{fig:GW150914}.
All the posteriors have confident support on the GR value, for all the events.
The BFs also favours GR, except for \GW{190828\_063405}\normalsize{} which has a slightly positive BF for non-GR and large variations in the deviations posteriors, especially in $\delta\tau_{22}$.
This result has already been observed in the LVK analysis~\cite{LIGOScientific:2020tif} with both the \texttt{pyRing} and \texttt{pSEOBNR} pipelines, where the overestimation of the damping time was tracked (through injections close to the trigger) to artifacts of noise fluctuations caused by the low SNR (in our analysis $\text{SNR}^{\text{net}}_{\text{opt}}\sim 5.0$).

\subsection{$M_f$-$a_f$ ``no-hair'' test}
\label{sec:Mf-af_TGR}

In this last section, we discuss an interesting representation of ``no-hair'' tests that can be constructed in the presence of at least two free modes.
This is inspired by GR tests on binary pulsars, see e.g. Fig.~6 of Ref.~\cite{Will:2014kxa}.
The achievable precision on such type of tests with future detectors upgrades has been studied for example in Ref.~\cite{Brito:2018rfr}.

\begin{figure*}
    \centering
    \hspace*{-0.8cm}\includegraphics[scale=1.35]{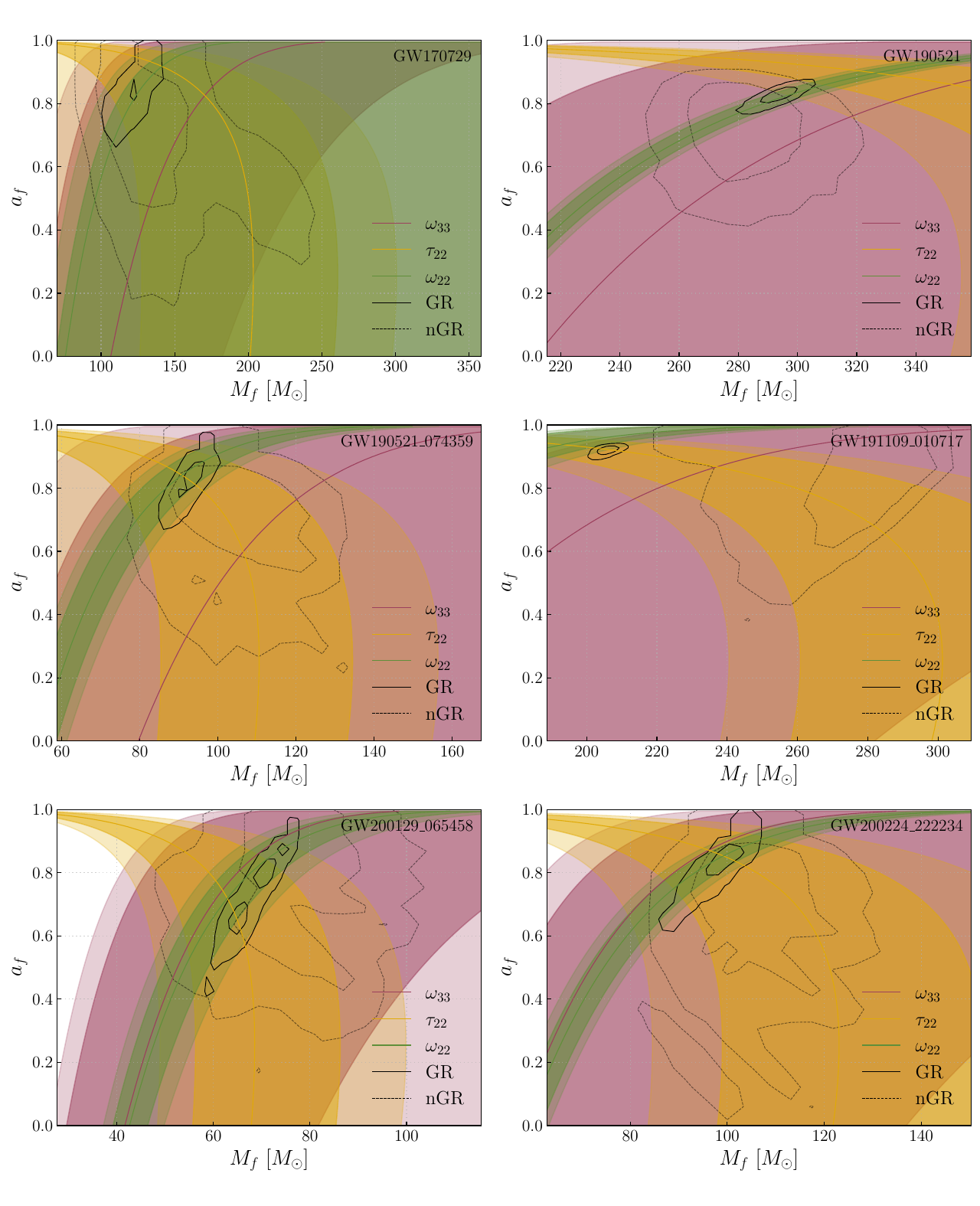}
    \captionsetup{format=plain, justification=raggedright, singlelinecheck=false}
    \caption{\footnotesize Projection of the non-GR effective QNM complex frequencies $\qty{\omega_{lm} = \omega_{lm}^{\text{\scriptsize{GR}}} (1 + \delta\omega_{lm}), \tau_{lm} = \tau_{lm}^{\text{\scriptsize{GR}}} (1 + \delta\tau_{lm})}$ in the plane of the final mass and spin $\qty{M_f, a_f}$.
    The \textit{green} bundle displays the median, $95\%$ and $84\%$ CIs of the projected samples from the fundamental mode effective frequency $\omega_{22}$, the \textit{pink} bundle from the effective frequency of the mode $(3,3)$, the \textit{yellow} bundle from the effective damping time of the fundamental mode $\tau_{22}$.
    The solid black curves represent the $50\%$ and $90\%$ credible levels of the corresponding GR analysis, i.e. the analysis with the same starting time but without GR deviations, and the dotted grey curves the same credible levels for the non-GR analysis.
    Each subplot refers to one of the events that showed evidence of the mode $(3,3)$.
    The samples used are taken from the analysis at the starting time corresponding with the maximum BF in favour of the mode $(3,3)$, as described in Sec. \ref{sec:Mf-af_TGR}.}
    \label{fig:TGR_plots}
\end{figure*}

We first compute the samples of the effective frequency using Eq.~\eqref{eq:effective_frequencies}, combining the GR and non-GR frequencies\footnote{For \texttt{TEOBPM}, this is done by first using the fits in~\cite{Jimenez-Forteza:2016oae} to compute $\qty{M_f, a_f}$ from $\qty{m_{1,2}, \chi_{1,2}}$, then the fits in~\cite{Berti:2005ys} to move from $\qty{M_f, a_f}$ to $\qty{f_{lm}, \tau_{lm}}$. Finally, the fractional deviations are summed through Eq.~\eqref{eq:effective_frequencies}.}.
We then project these values in the $\qty{M_f, a_f}$ plane.
Since there are several combinations of $\qty{M_f, a_f}$ which give the same frequency values, a sample of $f_{lm}$ is mapped onto a curve in the $\qty{M_f, a_f}$ plane.
Similarly, the set of all the $f_{22}$ samples, for example, generates a bundle in $\qty{M_f, a_f}$.
Repeating the same procedure with $f_{33}$ and $\tau_{22}$, we end up with three bundles of curves in the plane $\qty{M_f, a_f}$, as shown in Fig. \ref{fig:TGR_plots}.
At this point, two important considerations are due (see e.g.~\cite{Dreyer:2003bv, Berti:2005ys,Gossan:2011ha}).
\begin{itemize}
    \item If two frequencies and one damping time are detectable, the system presents three observables fixed by two parameters, $\qty{M_f, a_f}$. Thus, if the samples are consistent with GR, the three bundles should intersect in a single point in the high SNR limit, corresponding to the GR-predicted value of $\qty{M_f, a_f}$.
    If the observed effective frequencies and damping times are not consistent with GR, the three bundles will not intersect, as the two quantities $\qty{M_f, a_f}$ do not reproduce the three independent observables.
    \item A subtle step in the process is that the projection of the effective QNM frequencies is done assuming GR, in the sense that we invert the GR fits to obtain $\qty{M_f, a_f}$ from $\qty{f_{lm}, \tau_{lm}}$.
    This is crucial to bring up inconsistencies in the measured QNM frequencies assuming GR.
\end{itemize}
Fig. \ref{fig:TGR_plots} shows the outcome of this procedure for the subset of events with evidence for the mode $(3,3)$, together with the $50\%$ and $90\%$ CIs of the $\qty{M_f, a_f}$ GR and non-GR samples.
For each event, we use the posteriors from the analysis $\mathcal{H}_{\qty{\delta\omega_{22}, \delta\tau_{22}, \delta\omega_{33}}}$ at the starting time corresponding to the maximum BF for the mode $(3,3)$.
Specifically, we consider the samples obtained at $0M$ for \GW{170729}\normalsize, $-6M$ for \GW{190521}\normalsize, $-1M$ for \GW{190521\_074359}\normalsize, $-15M$ for \GW{191109\_010717}\normalsize, $3M$ for \GW{200129\_065458}\normalsize, $-2M$ for \GW{200224\_222234}\normalsize.
These arbitrary choices are motivated by an attempt to use a starting time that enhances the presence of the mode $(3,3)$, and could impact the test if some of the premerger signal is present in the data.
However, given the extremely broad bundles in Fig. \ref{fig:TGR_plots}, we believe that a detailed investigation at different starting times is not of interest.
Indeed, since the projected posteriors basically cover the entire $\qty{M_f, a_f}$ plane, we conclude that this test is still largely uninformative at current sensitivities, as expected from SNR estimates~\cite{Berti:2005ys,Baibhav:2023clw}.
Nevertheless, we highlight a few points.
First, \GW{200129\_065458}\normalsize{} is the loudest event, with $\text{SNR}^{\text{net}}_{\text{opt}}\sim 12.5$, and correspondingly has a tighter bundle, as expected.
With a perfect overlap between the GR and non-GR samples, and the projected $\omega_{33}$ not covering the entire plane, this event gives an idea of what a negative test will look like at higher sensitivities.
On the other hand, we point to the event \GW{191109\_010717}\normalsize{} with noise-induced GR deviations, to show how a non-GR analysis might manifest itself.
Here, we observe non-overlapping GR and non-GR posteriors, together with widely separated median curves in $\qty{\omega_{22}, \tau_{22}, \omega_{33}}$.
The projected $\omega_{22}$ in \GW{170729}\normalsize{} is much broader than in the other events because of its low $\text{SNR}^{\text{net}}_{\text{opt}}\sim 5.7$.
\GW{190521}\normalsize{} shows a different pattern from the other events, with the $\tau_{22}$ bundle outside the GR samples.
This behaviour can be traced in Fig. \ref{fig:GW190521}, where we observe a posterior in $\delta\tau_{22}$ that is shifted from zero at $t_0 = -6 M_f$.
This feature can be attributed to the systematics induced by the low-probability $t_{st}$, which was adopted to showcase the effect of systematics in this type of tests.

%==========================================================================
\section{Conclusions}
\label{sec:conclusions}

We presented a comprehensive ringdown analysis of the third catalog of gravitational-wave events GWTC-3 employing a highly-accurate template, \texttt{TEOBPM}, which increases the sensitivity of the search by modelling non-linearities in the early postmerger phase. 
Our work provides a robust framework for characterising the detectability of higher modes and deviations from general relativity.
With this method, we report a low-significance detection of the $(l,m) = (3,3)$ mode in four events, and tentative support for the Kerr hypothesis as predicted by general relativity (``no-hair'' tests) in the presence of two modes.

% Parameter estimation
The parameter estimation results with \texttt{TEOBPM} are in agreement with previous analyses from the LIGO-Virgo-KAGRA Collaboration. This, together with an increased precision in the constraints obtained compared to pure superposition of QNMs, serves to demonstrate the robustness and enhanced sensitivity of our framework compared to standard spectroscopic analyses.
Our results highlight the benefits of using ringdown models that include non-linearities, indicating how such models should play an important role in future ringdown analyses.

% HMs and TGR
The addition of the $(l,m)=(3,3)$ mode, albeit such mode is flagged with low-significance by Bayes Factors, still visibly affects the posterior distributions.
The fact that these results are observed for the first time in this work, for many events considered, is a combination of the improved waveform model used and of the systematic inclusion of the peak time uncertainty deployed in our analysis.
Beyond modes detection, such a robust assessment of start time-induced systematics is even more crucial when conducting tests of general relativity.
Assuming the presence of multiple modes, we allow for deviations in the QNM spectrum and find agreement with general relativistic predictions.

Our method could be further improved by repeating the analysis with inputs from different inspiral-merger-ringdown waveforms, gauging modeling systematics, and by characterising the statistical significance of the results on simulated signals, see e.g.~\cite{Maggio:2022hre}.
Future enhancements in the analysis methodology include the simultaneous marginalisation of sky position and start time together with the other analysis parameters.
It would also be interesting to combine the posteriors on the deviations to draw information on the population distribution.
This is the subject of ongoing studies.

% Conclusions
Our investigation will serve as a solid foundation for spectroscopic ringdown analyses with numerically informed models, both in the ongoing LIGO-Virgo-KAGRA observing run (O4) and with future more sensitive gravitational-waves detectors, nearing the possibility of robust multimodal tests of general relativity in the strong field regime.

%==========================================================================
\begin{acknowledgments}
V.G. thanks Stefano Rinaldi for invaluable assistance and suggestions, and for helping producing Fig. \ref{fig:TGR_plots}.
We thank Gunnar Riemenschneider for help in the implementation of TEOBResumSPM, and the TEOBResumS model developers for assistance, and for publicly maintaining the open source release \href{https://bitbucket.org/eob_ihes/teobresums/src/master/}{eob_ihes/teobresums}.
We thank Harrison Siegel for useful comments on the manuscript.
%
% NBI grants
V.G. and G.C. acknowledge support from the Villum Investigator program by the VILLUM Foundation (grant no. VIL37766) and the DNRF Chair program (grant no. DNRF162) by the Danish National Research Foundation.
V.G. acknowledges support form the French space agency CNES in the framework of LISA.
G.C. acknowledges support by the Della Riccia Foundation under an Early Career Scientist Fellowshipm, funding from the European Union’s Horizon 2020 research and innovation program under the Marie Sklodowska-Curie grant agreement No. 847523 ‘INTERACTIONS’.
This project has received funding from the European Union's Horizon 2020 research and innovation programme under the Marie Sklodowska-Curie grant agreement No 101131233.
%
% GWOSC
This research made use of data, software and/or web tools obtained from the
Gravitational Wave Open Science Center~\cite{LIGOScientific:2019lzm, KAGRA:2023pio}, a service
of the LIGO Scientific Collaboration, the KAGRA Collaboration and the Virgo
Collaboration.
%
% LVK Clusters
The authors are grateful for computational resources provided by the LIGO Laboratory (LHO) and supported by National Science Foundation Grants PHY-0757058 and PHY-0823459.
This material is based upon work supported by NSF's LIGO Laboratory which is a major facility fully funded by the National Science Foundation.
%
% LVK
LIGO Laboratory and Advanced LIGO are funded by the United States National Science Foundation (NSF) as well as the Science and Technology Facilities Council (STFC) of the United Kingdom, the Max-Planck Society (MPS), and the State of Niedersachsen/Germany for support of the construction of Advanced LIGO and construction and operation of the GEO600 detector. Additional support for Advanced LIGO was provided by the Australian Research Council. Virgo is funded, through the European Gravitational Observatory (EGO), by the French Centre National de Recherche Scientifique (CNRS), the Italian Istituto Nazionale di Fisica Nucleare (INFN) and the Dutch Nikhef, with contributions by institutions from Belgium, Germany, Greece, Hungary, Ireland, Japan, Monaco, Poland, Portugal, Spain. KAGRA is supported by Ministry of Education, Culture, Sports, Science and Technology (MEXT), Japan Society for the Promotion of Science (JSPS) in Japan; National Research Foundation (NRF) and Ministry of Science and ICT (MSIT) in Korea; Academia Sinica (AS) and National Science and Technology Council (NSTC) in Taiwan.\\

\noindent{\textbf{\textit{Software}}.} The violin plots and the ridgeline plots have been produced using the open source \texttt{python} package \texttt{namib}~\cite{namib}, which makes use of the \texttt{statsmodels}~\cite{seabold2010statsmodels} and \texttt{joypy}~\cite{joypy} packages.
The manuscript content has been derived using publicly available software: \texttt{matplotlib, numpy, pyRing, raynest, scipy, seaborn, sxs}~\cite{matplotlib, numpy, pyRing, raynest, scipy, seaborn, sxs}.

%----------------------------------------------------------------------------------------

\end{acknowledgments}

%%%%%%%%%%%%%%%%%%%%%%%%%%%%%%%%%%%%%%%%%%%%%%%%%%%%%%%%%%%%%%%%%%%%%%%%%%%%%%%
% APPENDIX
\newpage
\appendix
\section*{Appendix}
%==========================================================================
\subsection{Geometric units in \texttt{TEOBPM}}
\label{sec:appendix_a}

In \texttt{TEOBPM}, the expansion coefficients $h_{lm}$ are rescaled through the following transformations.
\begin{itemize}
    \item Define the \textit{mass-distance units conversion factor} as
    \begin{equation}
        C_{md} \equiv \frac{\SI{}{M_{\odot}}}{\SI{}{\mega\parsec}}\frac{G}{c^2},\quad\quad [C_{md}] = 1,
    \end{equation}
    which converts a mass expressed in solar masses $\SI{}{M_{\odot}}$ into a distance in Megaparsesc ($\SI{}{\mega\parsec}$).
    The waveform amplitude is rescaled by
    \begin{equation}
        h_{lm} \longrightarrow \frac{r}{\nu M C_{md}}\,h_{lm},
    \end{equation}
    where $M\equiv m_1+m_2$ is the total mass of the binary, and $\nu\equiv \frac{m_1 m_2}{(m_1+m_2)^2}$ is the symmetric mass ratio (this convention is specific to \texttt{TEOBPM} and e.g. NR waveforms do not follow this convention).
    \item Define the \textit{mass-time units conversion factor} as
    \begin{equation}
        C_{mt} \equiv \SI{}{M_{\odot}}\frac{G}{c^3},\quad\quad [C_{mt}]=\SI{}{s},
    \end{equation}
    which converts a mass expressed in solar masses $\SI{}{M_{\odot}}$ into a time in seconds.
    The time variable in expressed in units of $M_f$, i.e. the black hole final mass expressed in solar masses,
    \begin{equation}\label{mass_time_convertion}
        t \longrightarrow \tau \equiv \frac{t-t_{\text{mrg}}}{M_f C_{mt}},
    \end{equation}
    where $t_{\text{mrg}}$ is the \textit{merger time}, defined as the time at which the maximum of the waveform's amplitude of the fundamental mode $(2,2)$ occurs, that is $t_{\text{mrg}}\equiv t^{\text{peak}}_{22}$, corresponding to the maximum of $|h_{22}(t)|$.
\end{itemize}
With the above transformations, the waveform is said to be expressed in \textit{geometric units}.
Adhering to the conventions used in the rest of the text, we set $G=c=\SI{}{M_{\odot}}=1$, and the conversion factors become $C_{md}=1/\SI{}{\mega\parsec}$ and $C_{mt}=1$.
Note that at the end of the implementation, these transformations need to be reversed to obtain the properly scaled signal in physical units.

%==========================================================================
\subsection{Quality assessment of \texttt{TEOBPM}}
\label{sec:appendix_b}

We ascertain the accuracy of the \texttt{TEOBPM} model against numerical relativity simulations from the \textit{Simulating eXtreme Spacetimes catalog} (SXS) of numerical simulations of merging black holes \cite{Mroue:2013xna, Boyle:2019kee}.
Specifically, we have selected the simulations with spin aligned or anti-aligned through the \texttt{sxs} package~\cite{sxs}, and isolated the ringdown part cutting the waveform at the time corresponding to the peak of each mode. 
To compare the simulations with our implementation of \texttt{TEOBPM} in \texttt{pyRing}, we generate the \texttt{TEOBPM} waveform from the parameters of the NR simulation at \textit{reference time}, for each mode separately.
The initial phase of the \texttt{TEOBPM} waveform is chosen to maximise the match with the simulation.
We use the \textit{mismatch} (MM) to quantify the agreement between the two waveforms,
\begin{equation}\label{mismatch_fitting_factor}
    \text{FF}\equiv \frac{\vb{a}\cdot\vb{b}}{\sqrt{(\vb{a}\cdot\vb{a})(\vb{b}\cdot\vb{b})}},\quad \text{MM}\equiv 1-\text{FF},
\end{equation}
where FF is the \textit{fitting factor}, and the time domain scalar product is defined as
\begin{equation}\label{time_domain_scalar_product}
    \vb{a}\cdot\vb{b} \equiv \abs{\sum_i a_i \bar{b}_i} \equiv \abs{\vb{a}^T\bar{\vb{b}}}.
\end{equation}
The bar indicates complex conjugation and the ${}_T$ transposition.
Note that, in the definition of FF, there is no need to maximise the scalar product over the parameter space since the two waveforms are the closest possible by construction.
The mismatch takes values in the interval $[0,1]$.

For the modes $\qty{(2,2),(2,1),(3,3),(3,2),(4,4),(4,3)}$, Fig. \ref{mismatch_catalog_h_fig} shows the percent mismatch of the scaled expansion coefficients $r(h_{lm})/M$, as function of the mass ratio $q$ and the final adimensional spin $a_f$.
The quantities $h_{lm}$ are defined in Eq.~\eqref{rescaled_teob_wf_complex}.
Each dot in the figure corresponds to a simulation, for a total of 578 simulations.
The various modes show different accuracy.
\begin{itemize}
    \item The fundamental mode is in general very accurate, with a mean percent mismatch of $\text{MM}\% = 0.008 \pm 0.009$.
    
    \item The mode $(3,3)$ is also highly reliable, with a mean of $\text{MM}\% = 1.0 \pm 2.0$.
    We note, however, that the largest mismatches occur for equal-mass systems $q\sim 1$, for which this mode is suppressed by symmetry.
    This suggests that the higher values of the mismatch in such simulations is guided by numerical uncertainties.
    
    \item The mode $(2,1)$ has mismatches comparable with those of the mode $(3,3)$, $\text{MM}\% = 2.0 \pm 9.0$, except for a subset of simulations in which the accuracy drops.
    These simulations lie in a region of the parameter space in which this mode is not suppressed.
    
    \item The remaining modes $\qty{(3,2), (4,4), (4,3), (4,2)}$ show in general larger values of the mismatch, typically around $10\%$ and even higher for some modes (although the mode $(4,4)$ tends to perform better than the others).
    This is not of great concern, since these are subleading modes, and contribute very little to the overall GW signal.
    Given their mismatch values and total contribution to the waveform, we decide to not use these modes in this work, noting that improved fits for these modes will be needed in the future, when performing analyses at higher sensitivities.
\end{itemize}
Eventually, we want to stress an important fact, that the mismatches discussed above are computed on the different modes \textit{separately}, and do not refer to the waveform given by the sum of the various modes.
Consequently, we are missing a complementary information given by the relative excitation between the fundamental mode and the different HMs.
Since the amplitude of the HMs is only a small fraction of that of the the fundamental mode for the systems considered, the mismatch on the \textit{global} waveform is in general very close to that of the mode $(2,2)$ alone, although this aspect has not been studied systematically.

%==========================================================================
\subsection{SNR in time domain}
\label{sec:appendix_c}

% Experimental setup
The output of interferometric measurements can be described as
\begin{equation}\label{data}
    d(t) = h(t,\vb*{\theta}) + n(t),
\end{equation}
where $h(t,\vb*{\theta})$ is a deterministic model of the observables, embedded in some instrumental noise $n(t)$.
Note the explicit dependence of $h(t,\vb*{\theta})$ on a set of parameters $\vb*{\theta}$ that parametrize the model.
The noise $n(t)$ is treated as a stochastic variable.
More in detail, the measurements consist of discrete \textit{time series} of the sampled signal $d(t)$ at $n$ different times, such that
\begin{equation}\label{data_sampled}
    \vb{d}=\vb{h}+\vb{n},\quad\text{with}\quad \vb{d}\equiv\{d(t_1),\dots,d(t_n)\},
\end{equation}
and the same goes for $\vb{h}$ and $\vb{n}$.
%
%
% SNR definition
We define the \textit{signal-to-noise ratio} (SNR) as the ratio between the \textit{power} of the model $\mathbb{P}_{\vb{h}}$ and that of the noise $\mathbb{P}_{\vb{n}}$,
\begin{equation}\label{snr_general}
    \text{SNR} \equiv \frac{\mathbb{P}_{\vb{h}}}{\mathbb{P}_{\vb{n}}} \equiv \frac{E[\vb{h}^T\vb{h}]}{E[\vb{n}^T\vb{n}]} =  \frac{\vb{h}^T\vb{h}}{\mathbb{C}_{\vb{n}\vb{n}}},
\end{equation}
where the power is the expectation value of squared time series, and we used the fact that $\vb{h}$ is deterministic.
In the last passage we introduce the \textit{autocovariance matrix} of the noise.

% SNR derivation
We now outline a procedure to re-weight $\vb{h}$ and $\vb{n}$ in such a way that the SNR is maximised.
We introduce the \textit{linear filter} $\vb{k}$, (for the moment an unknown time series), that can be applied to a vector $\vb{a}$ through the standard scalar product
\begin{equation}\label{scalar_product_td}
    \hat{\vb{a}} \equiv \sum_i a_i k_i \equiv \vb{a}^T\vb{k}.
\end{equation}
To find the filter $\vb{k}$ that maximises the SNR~\eqref{snr_general}, we rewrite the latter  in terms of the filtered functions $\hat{\vb{h}}$ and $\hat{\vb{n}}$
\begin{multline}\label{snr_opt_1}
    \widehat{\text{SNR}} = \frac{\mathbb{P}_{\hat{\vb{h}}}}{\mathbb{P}_{\hat{\vb{n}}}} = \frac{E[(\vb{h}^T\vb{k})^T(\vb{h}^T\vb{k})]}{E[(\vb{n}^T\vb{k})^T(\vb{n}^T\vb{k})]}\\
    = \frac{E[\qty{\vb{k}^T \vb{h}}^2]}{E[\vb{k}^T\vb{n}\vb{n}^T\vb{k}]} = \frac{\qty{\vb{k}^T \vb{h}}^2}{\vb{k}^TE[\vb{n}^T\vb{n}]\vb{k}} = \frac{\qty{\vb{k}^T \vb{h}}^2}{\vb{k}^T\mathbb{C}_{\vb{n}\vb{n}}\vb{k}},
\end{multline}
where we have used that also the filter $\vb{k}$ is deterministic.
Define the scalar product between two vectors $\vb{a}$ and $\vb{b}$, weighted with the inverse autocovariance matrix
\begin{equation}\label{scalar_product_cov}
    \bra{\vb{a}}\ket{\vb{b}} \equiv \vb{a}^T \mathbb{C}^{-1}_{\vb{n}\vb{n}} \vb{b}.
\end{equation}
Then, Eq.~\eqref{snr_opt_1} can be written in terms of the scalar product \eqref{scalar_product_cov} as
\begin{equation}\label{snr_opt_2}
    \widehat{\text{SNR}} = \frac{\bra{\vb{\mathbb{C}_{\vb{n}\vb{n}}\vb{k}}}\ket{\vb{h}}^2}{\bra{\vb{\mathbb{C}_{\vb{n}\vb{n}}\vb{k}}}\ket{\vb{\mathbb{C}_{\vb{n}\vb{n}}\vb{k}}}} = \frac{\qty{(\mathbb{C}_{\vb{n}\vb{n}}\vb{k})^T \mathbb{C}^{-1}_{\vb{n}\vb{n}} \vb{h}}^2}{(\mathbb{C}_{\vb{n}\vb{n}}\vb{k})^T \mathbb{C}^{-1}_{\vb{n}\vb{n}} \mathbb{C}_{\vb{n}\vb{n}}\vb{k}} = \frac{\qty{\vb{k}^T \vb{h}}^2}{\vb{k}^T\mathbb{C}_{\vb{n}\vb{n}}\vb{k}}.
\end{equation}
Eq.~\eqref{snr_opt_2} conveys that the filtered SNR is proportional to the square of the scalar product $\bra{\vb{\mathbb{C}_{\vb{n}\vb{n}}\vb{k}}}\ket{\vb{h}}$, therefore it is maximum when this scalar product is maximum.
Geometrically, this happens when the vector $\mathbb{C}_{\vb{n}\vb{n}}\vb{k}$ is parallel to $\vb{h}$.
Therefore, the filter $\vb{k}^*$ that maximises the SNR is
\begin{equation}\label{matched_filter}
    \mathbb{C}_{\vb{n}\vb{n}}\vb{k}^* \parallel \vb{h}\quad\Longrightarrow\quad \vb{k}^*\equiv \alpha\: \mathbb{C}^{-1}_{\vb{n}\vb{n}}\vb{h},
\end{equation}
for some constant $\alpha$.
$\vb{k}^*$ is known as the \textit{matched filter}.
Substituting the matched filter \eqref{matched_filter} in Eq.~\eqref{snr_opt_1}, we obtain the \textit{optimal signal-to-noise ratio}
\begin{multline}\label{optimal_snr}
    \text{SNR}_{\text{opt}} \equiv \frac{\qty{\vb{k}^{*T} \vb{h}}^2}{\vb{k}^{*T}\mathbb{C}_{\vb{n}\vb{n}}\vb{k}^*} = \frac{\alpha^2\qty{(\mathbb{C}^{-1}_{\vb{n}\vb{n}} \vb{h})^T\vb{h}}^2}{\alpha^2(\mathbb{C}^{-1}_{\vb{n}\vb{n}} \vb{h})^T \mathbb{C}_{\vb{n}\vb{n}} (\mathbb{C}^{-1}_{\vb{n}\vb{n}} \vb{h})}\\
    = \frac{\qty{\vb{h}^T \mathbb{C}^{-1}_{\vb{n}\vb{n}} \vb{h}}^2}{\vb{h}^T \mathbb{C}^{-1}_{\vb{n}\vb{n}} \vb{h}} = \frac{\bra{\vb{h}}\ket{\vb{h}}^2}{\bra{\vb{h}}\ket{\vb{h}}} = \bra{\vb{h}}\ket{\vb{h}}.
\end{multline}
In analogy with the optimal signal-to-noise ratio $\text{SNR}_{\text{opt}}$, we also define the \textit{matched filter signal-to-noise ratio} as
\begin{equation}\label{matched_filter_snr}
    \text{SNR}_{\text{mf}} \equiv \frac{\bra{\vb{h}}\ket{\vb{d}}}{\sqrt{\bra{\vb{h}}\ket{\vb{h}}}}.
\end{equation}

%==========================================================================
\subsection{Analytical prediction of HMs}
\label{sec:appendix_d}

Given two competing hypotheses $\mathcal{H}_{lm}$ and $\mathcal{H}_{22}$, we define the \textit{Bayes factor} $\mathcal{B}_{lm,22}$
\begin{equation}\label{odds_ratio}
    \frac{p(\mathcal{H}_{lm}|\vb{d},I)}{p(\mathcal{H}_{22}|\vb{d},I)} = \frac{p(\mathcal{H}_{lm}|I)}{p(\mathcal{H}_{22}|I)} \frac{p(\vb{d}|\mathcal{H}_{lm},I)}{p(\vb{d}|\mathcal{H}_{22},I)} \equiv \frac{p(\mathcal{H}_{lm}|I)}{p(\mathcal{H}_{22}|I)}\: \mathcal{B}_{{lm}.{22}},
\end{equation}
If $\mathcal{H}_i$ is parametrized by $\vb*{\theta}_i \in \Theta_i$, from Bayes theorem follows that the likelihood $p(\vb{d}|\mathcal{H}_i,I)$ is
\begin{equation}\label{likelihood_ockham_factor}
    p(\vb{d}|\mathcal{H}_i,I) = \int_{\Theta_i} p(\vb{d}|\vb*{\theta}_i,\mathcal{H}_i,I)\:p(\vb*{\theta}_i|\mathcal{H}_i,I)  \dd{\vb*{\theta}_i} \equiv \mathcal{W}_i\,L_{\text{max},i},
\end{equation}
In Eq.~\eqref{likelihood_ockham_factor} we have introduced the \textit{Ockham factor} $\mathcal{W}$ and the maximum likelihood value $L_{\text{max}}$, defined as
\begin{flalign}\label{ockham_factor_def}
    \nonumber
    \mathcal{W}_i \equiv \frac{1}{L_{\text{max},i}} \int_{\Theta_i} p(\vb{d}|\vb*{\theta}_i,\mathcal{H}_i,I)\:p(\vb*{\theta}_i|\mathcal{H}_i,I)  \dd{\vb*{\theta}_i},\\
    L_{\text{max},i} \equiv \text{max}\qty{p(\vb{d}|\vb*{\theta}_i,\mathcal{H}_i,I)}.
\end{flalign}
Setting the prior ratio equal to one and taking the logarithm, Eq.~\eqref{odds_ratio} becomes
\begin{equation}\label{bayes_factor_hms_init}
    \ln \mathcal{B}_{lm,22} = \ln{\mathcal{L}_{\text{max},lm}} - \ln{\mathcal{L}_{\text{max},22}} + \ln\qty{\frac{\mathcal{W}_{lm}}{\mathcal{W}_{22}}}.
\end{equation}

The likelihood of a deterministic model can be written as
\begin{multline}
    \ln \mathcal{L} = -\frac{1}{2}\braket{\vb{d}-\vb{h}}{\vb{d}-\vb{h}}\\ = -\frac{1}{2}\bra{\vb{d}}\ket{\vb{d}} + \bra{\vb{d}}\ket{\vb{h}} -\frac{1}{2}\bra{\vb{h}}\ket{\vb{h}} +\text{const.},
\end{multline}
where we have used the fact that the scalar product in Eq.~\eqref{scalar_product_cov} is symmetric.
This result can be derived from the maximum entropy principle, as proven in Sec. 4.5 and App. B.2 of \cite{Gennari:MScThesis}.
Note that the data $\vb{d}$ are fixed, whereas the waveform depends on the set of parameters $\vb*{\theta}\in\Theta_{lm}$ introduced in section \eqref{subsec:TEOBPM_characteristics}.
Thus, the scalar product $\braket{\vb{d}}{\vb{h}}$ varies over the parameter space $\Theta_{lm}$.
Consider a rescaling of the waveform $\vb{h}$ by some constant $\alpha$, then
\begin{multline}\label{log_likelihood_alpha}
    \vb{h}\to\alpha\vb{h}\\ \quad\Longrightarrow\quad \ln{\mathcal{L}}\to -\frac{1}{2}\bra{\vb{d}}\ket{\vb{d}} + \alpha\bra{\vb{d}}\ket{\vb{h}} -\frac{\alpha^2}{2}\bra{\vb{h}}\ket{\vb{h}} +\text{const.},
\end{multline}
where we used the linearity of the scalar product.
The maximum of the logarithmic likelihood with respect to $\alpha$ is given by
\begin{equation}
    0 = \eval{\dv{\alpha}\ln \mathcal{L}}_{\alpha=\alpha^*} = \bra{\vb{d}}\ket{\vb{h}} - \alpha^*\bra{\vb{h}}\ket{\vb{h}} \quad\Longrightarrow\quad \alpha^* = \frac{\bra{\vb{d}}\ket{\vb{h}}}{\bra{\vb{h}}\ket{\vb{h}}},
\end{equation}
and from Eq.~\eqref{log_likelihood_alpha} it follows that
\begin{multline}\label{log_likelihood_alpha_2}
    \eval{\ln \mathcal{L}}_{\alpha=\alpha^*} = -\frac{1}{2}\bra{\vb{d}}\ket{\vb{d}} + \alpha^{*2}\bra{\vb{h}}\ket{\vb{h}} - \frac{\alpha^{*2}}{2}\bra{\vb{h}}\ket{\vb{h}} +\text{const.}\\
    = \frac{1}{2}\qty{\alpha^{*2}\bra{\vb{h}}\ket{\vb{h}} - \bra{\vb{d}}\ket{\vb{d}}} = \frac{1}{2}\qty{\frac{\braket{\vb{d}}{\vb{h}}^2}{\braket{\vb{h}}{\vb{h}}} - \braket{\vb{d}}{\vb{d}}} +\text{const.}.
\end{multline}

By definition, the maximum logarithmic likelihood is given by
\begin{multline}\label{bayes_factor_hms_temp}
    \ln{\mathcal{L}_{\text{max}}} = \max_{\vb*{\theta}\in\Theta_{lm}} \qty{\frac{1}{2}\frac{\braket{\vb{d}}{\vb{h}}^2}{\braket{\vb{h}}{\vb{h}}}} - \frac{1}{2}\braket{\vb{d}}{\vb{d}},\\
    \Longrightarrow\quad 2\qty{\ln{\mathcal{L}_{\text{max},lm}} - \ln{\mathcal{L}_{\text{max},22}}}\\
    = \max_{\vb*{\theta}\in\Theta_{lm}} \qty{\frac{\braket{\vb{d}}{\vb{h}_{lm}}^2}{\braket{\vb{h}_{lm}}{\vb{h}_{lm}}} - \frac{\braket{\vb{d}}{\vb{h}_{22}}^2}{\braket{\vb{h}_{22}}{\vb{h}_{22}}}},
\end{multline}
having assumed that the parameter space of the hypothesis $\mathcal{H}_{22}$ is nested in that of $\mathcal{H}_{lm}$, so that $\Theta_{22}\subseteq \Theta_{lm}$\footnote{This is the usual scenario in which one hypothesis is an extension of the other. For example, in \texttt{TEOBPM}, the inclusion of one higher mode in the analysis comes with an additional degree of freedom, the mode phase at merger, as discussed in section \ref{subsec:TEOBPM_characteristics}}.
Assuming that the waveform $\vb{h}_{lm}$ which maximises $\mathcal{L}_{lm}$ is close to the data, we can expand the data as
\begin{equation}\label{wf_expansion_equation}
    \vb{d} \equiv \vb{h}_{lm} + \Delta \vb{h},\quad\Longrightarrow\quad \braket{\vb{d}}{\vb{h}} = \braket{\vb{h}_{lm}}{\vb{h}} + \mathcal{O}\qty{\Delta\vb{h}}.
\end{equation}
Then, Eq.~\eqref{bayes_factor_hms_temp} becomes
\begin{multline}\label{Ockham_factor_equation}
    \max_{\vb*{\theta}\in\Theta_{lm}} \qty{\frac{\braket{\vb{d}}{\vb{h}_{lm}}^2}{\braket{\vb{h}_{lm}}{\vb{h}_{lm}}} - \frac{\braket{\vb{d}}{\vb{h}_{22}}^2}{\braket{\vb{h}_{22}}{\vb{h}_{22}}}}\\
    = \max_{\vb*{\theta}\in\Theta_{lm}} \qty{\braket{\vb{h}_{lm}}{\vb{h}_{lm}} - \frac{\braket{\vb{h}_{lm}}{\vb{h}_{22}}^2}{\braket{\vb{h}_{22}}{\vb{h}_{22}}}}  + \mathcal{O}\qty{\Delta\vb{h}}\\
    = \max_{\vb*{\theta}\in\Theta_{lm}} \qty{\qty(1 - \frac{\braket{\vb{h}_{lm}}{\vb{h}_{22}}^2}{\braket{\vb{h}_{lm}}{\vb{h}_{lm}}\braket{\vb{h}_{22}}{\vb{h}_{22}}}) \braket{\vb{h}_{lm}}{\vb{h}_{lm}}}\\
    + \mathcal{O}\qty{\Delta\vb{h}} = \qty(1-\text{FF}^2)\, \text{SNR}_{\text{opt}}^2 + \mathcal{O}\qty{\Delta\vb{h}},
\end{multline}
where in the last line we used the definitions of the FF \eqref{eq:fitting_factor} and optimal SNR \eqref{optimal_snr}.
Using equation \eqref{bayes_factor_hms_init}, we finally arrive at the relation
\begin{equation}
    \ln{\mathcal{B}_{lm,22}} = \frac{1}{2}\qty(1-\text{FF}^2)\, \text{SNR}_{\text{opt}}^2 + \ln\qty{\frac{\mathcal{W}_{lm}}{\mathcal{W}_{22}}} + \mathcal{O}\qty{\Delta\vb{h}}.
\end{equation}

%==========================================================================
\subsection{Detectability of the modes $(3,2)$ and $(4,4)$}
\label{sec:appendix_e}

\begin{figure*}
    \hspace*{-0.5cm}\includegraphics[scale=0.84]{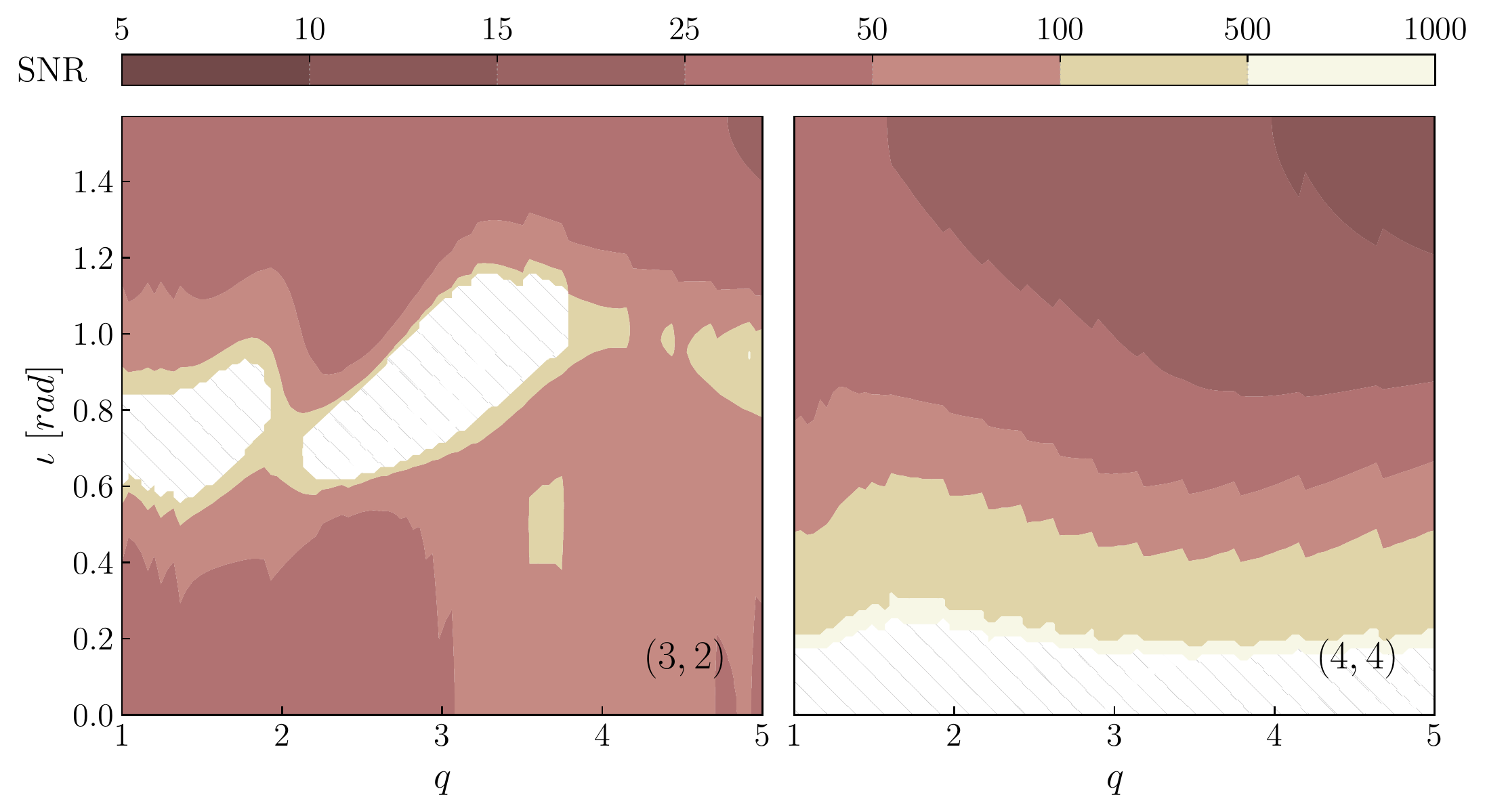}
    \caption{\footnotesize Colormap of the optimal signal-to-noise ratio needed to measure a logarithmic Bayes factor of $\ln{\mathcal{B}_{lm,22}}=3$, in favour of the mode $(3,2)$ (left) and $(4,4)$ (right) over the fundamental mode.
    The results are for two nonspinning initial black holes $\chi_1=\chi_2=0$, and are expressed as function of the mass ratio $q$ and inclination $\iota$.
    The dashed regions correspond to $\text{SNR}>1000$.}
    \label{fig:snr_32-44}
\end{figure*}

Fig. \ref{fig:snr_32-44} shows the prediction on the detectability for the modes $\qty{(3,2), (4,4)}$ as a function of the mass ratio $q$ and inclination $\iota$.
The threshold for detection is arbitrarily set as $\ln{\mathcal{B}_{lm,22}}=3$; in colours, the optimal SNR.
Compared to Fig. \ref{fig:snr_33-21}, which shows the same results for the dominant modes $\qty{(3,3), (2,1)}$, note the increase in the SNR required to detect the modes $\qty{(3,2), (4,4)}$.
Indeed, at current sensitivities in the RD, $\text{SNR}\sim 15-25$, only the mode $(4,4)$ could be detected for both large inclinations and mass ratios.
Finally, we note how the detectability pattern of the mode $(3,2)$ highly differs from that of the other modes. This is expected due to the large contributions received by mode-mixing~\cite{Berti:2005gp,Buonanno:2006ui,Kelly:2012nd,Berti:2014fga,London:2018nxs}.

%==========================================================================
\subsection{Input parameters of \texttt{pyRing}}
\label{sec:appendix_f}
\twocolumngrid

In table \ref{tab:input_parameters_pyring_table}, we report the values of the input parameters from IMR used to perform our RD analysis with \texttt{pyRing}.
The starting time $t_0$ is estimated from the peak of preferred IMR analysis by the LVK collaboration~\cite{LIGOScientific:2021djp}. 
Specifically, for each set of parameters samples from the IMR analysis we generate the corresponding signal and compute the time at which the modulus of $h_{+}^2 + h_{\times}^2$ occurs. The starting time is selected as the median value of the constructed distribution.
The time is expressed in the time scale Global Positioning System (GPS) and is referred to the LIGO Hanford detector.
If LIGO Hanford was offline, $t_0$ refers to LIGO Livingstone.
Also the sky position values are estimated from the IMR analysis of the LVK Collaboration.
Specifically, the values in table \ref{tab:input_parameters_pyring_table} are the median values of the posterior distributions for the following models:
\begin{itemize}
    \item \texttt{SEOBNRv4ROM} for O1 and O2;
    \item \texttt{IMRPhenomPv2} for O3a, except for
    \begin{itemize}
        \item \texttt{IMRPhenomPv3HM} for \GW{190412}\normalsize;
        \item \texttt{NRSur7dq4} for \GW{190521}\normalsize;
        \item \texttt{IMRPhenomPv3HM} for \GW{190814}\normalsize;
    \end{itemize}
    \item \texttt{IMRPhenomXPHM} for O3b, except for \texttt{IMRPhenomPv2} for \GW{191215\_223052}\normalsize.
\end{itemize}
For our RD analysis, we used cleaned data publicly available and downloaded from the GWOSC online database.
Specifically, they have been taken from the channels \texttt{H1\textunderscore HOFT\textunderscore CLEAN\textunderscore SUB60HZ\textunderscore C01}, \texttt{L1\textunderscore HOFT\textunderscore CLEAN\textunderscore SUB60HZ\textunderscore C01}, \texttt{V1Online}, as clarified at the end of the webpage \url{https://www.gw-openscience.org/GWTC-3/}.

%%%%%%%%%%%%%%%%%%%%%%%%%%%%%%%%%%%%%%%%%%%%%%%%%%%%%%%%%%%%%%%%%%%%%%%%%%%%%%%
% ADDITIONAL TABLES AND FIGURES

% Table HMs
\setlength{\tabcolsep}{8pt}
\begin{table*}

    \centering
    \begin{tabular}{ L{4cm} R{0.8cm} R{0.8cm} C{0.8cm} R{1.1cm} C{0.5cm} R{0.8cm} R{0.8cm} C{0.8cm} R{1.2cm} }
        \toprule\toprule
        & \multicolumn{4}{c}{$lm$$:\qty{22}$} & & \multicolumn{4}{c}{$lm$$:\qty{22,33}$} \\
        \cmidrule(lr){2-5} \cmidrule(lr){7-10} \noalign{\vskip-9pt}
        \bf{event} & $\text{SNR}^{\text{net}}_{\text{opt}}$ & $H$ & $\text{lnL}_{\text{max}}$ & $\ln{\mathcal{B}_{lm,22}}$ & & $\text{SNR}^{\text{net}}_{\text{opt}}$ & $H$ & $\text{lnL}_{\text{max}}$ & $\ln{\mathcal{B}_{lm,22}}$    \\
        
        \midrule

        &&&&&&&&& \\[-6mm]
        \GW{150914}         & $12.5_{-1.3}^{+1.3}$ & $14.4$ & \scriptsize{$38775$} & / & & $12.5_{-1.3}^{+1.3}$ & $14.6$ & \scriptsize{$38775$} & $-0.26$ \\[-2mm]
        \GW{170729}         & $ 5.1_{-1.4}^{+1.4}$ & $ 7.7$ & \scriptsize{$57689$} & / & & $ 5.1_{-1.4}^{+1.4}$ & $ 7.8$ & \scriptsize{$57689$} & $ 0.03$ \\[-2mm]
        \GW{170823}         & $ 6.2_{-1.3}^{+1.3}$ & $ 8.9$ & \scriptsize{$38951$} & / & & $ 6.2_{-1.3}^{+1.3}$ & $ 9.0$ & \scriptsize{$38952$} & $-0.12$ \\[-2mm]
        
        \GW{190503\_185404} & $ 5.7_{-1.4}^{+1.3}$ & $ 8.8$ & \scriptsize{$58528$} & / & & $ 5.7_{-1.4}^{+1.4}$ & $ 8.9$ & \scriptsize{$58529$} & $-0.31$ \\[-2mm]
        \GW{190519\_153544} & $ 9.5_{-1.3}^{+1.3}$ & $ 9.4$ & \scriptsize{$58603$} & / & & $ 9.5_{-1.3}^{+1.3}$ & $ 9.7$ & \scriptsize{$58604$} & $-0.15$ \\[-2mm]
        \GW{190521}         & $12.1_{-1.3}^{+1.3}$ & $ 9.9$ & \scriptsize{$58560$} & / & & $12.1_{-1.3}^{+1.3}$ & $10.2$ & \scriptsize{$58560$} & $-0.30$ \\[-2mm]
        \GW{190521\_074359} & $ 9.5_{-1.3}^{+1.3}$ & $11.8$ & \scriptsize{$39411$} & / & & $ 9.6_{-1.3}^{+1.3}$ & $11.9$ & \scriptsize{$39413$} & $ 0.22$ \\[-2mm]
        \GW{190602\_175927} & $ 8.7_{-1.3}^{+1.3}$ & $ 9.2$ & \scriptsize{$58544$} & / & & $ 8.7_{-1.3}^{+1.3}$ & $ 9.5$ & \scriptsize{$58545$} & $-0.35$ \\[-2mm]
        \GW{190706\_222641} & $ 7.8_{-1.4}^{+1.3}$ & $10.9$ & \scriptsize{$58456$} & / & & $ 7.8_{-1.4}^{+1.3}$ & $11.0$ & \scriptsize{$58457$} & $-0.20$ \\[-2mm]
        \GW{190727\_060333} & $ 5.8_{-1.4}^{+1.4}$ & $ 8.2$ & \scriptsize{$58577$} & / & & $ 5.8_{-1.4}^{+1.4}$ & $ 8.2$ & \scriptsize{$58579$} & $ 0.10$ \\[-2mm]
        \GW{190828\_063405} & $ 5.0_{-1.5}^{+1.4}$ & $ 9.0$ & \scriptsize{$58570$} & / & & $ 5.0_{-1.4}^{+1.4}$ & $ 9.0$ & \scriptsize{$58570$} & $ 0.01$ \\[-2mm]
        \GW{190910\_112807} & $ 8.1_{-1.3}^{+1.3}$ & $10.6$ & \scriptsize{$38854$} & / & & $ 8.1_{-1.3}^{+1.3}$ & $10.7$ & \scriptsize{$38855$} & $-0.13$ \\[-2mm]
        \GW{190915\_235702} & $ 6.2_{-1.4}^{+1.3}$ & $10.4$ & \scriptsize{$58438$} & / & & $ 6.3_{-1.4}^{+1.3}$ & $10.7$ & \scriptsize{$58440$} & $-0.21$ \\[-2mm]
        
        \GW{191109\_010717} & $11.9_{-1.3}^{+1.3}$ & $15.1$ & \scriptsize{$39407$} & / & & $12.0_{-1.3}^{+1.3}$ & $15.5$ & \scriptsize{$39408$} & $ 0.20$ \\[-2mm]
        \GW{191222\_033537} & $ 5.6_{-1.4}^{+1.3}$ & $7.0$  & \scriptsize{$39418$} & / & & $ 5.6_{-1.3}^{+1.3}$ & $ 7.2$ & \scriptsize{$39418$} & $-0.11$ \\[-2mm]
        \GW{200129\_065458} & $12.4_{-1.3}^{+1.3}$ & $13.5$ & \scriptsize{$58419$} & / & & $12.4_{-1.3}^{+1.3}$ & $13.6$ & \scriptsize{$58420$} & $ 0.17$ \\[-2mm]
        \GW{200224\_222234} & $10.2_{-1.3}^{+1.3}$ & $11.6$ & \scriptsize{$58548$} & / & & $10.2_{-1.3}^{+1.3}$ & $11.7$ & \scriptsize{$58549$} & $ 0.05$ \\[-2mm]
        \GW{200311\_115853} & $ 7.1_{-1.4}^{+1.3}$ & $11.2$ & \scriptsize{$58549$} & / & & $ 7.0_{-1.3}^{+1.3}$ & $11.2$ & \scriptsize{$58550$} & $-0.03$ \\[1mm]

        \midrule
        
        & \multicolumn{4}{c}{$lm$$:\qty{22,21}$} & & \multicolumn{4}{c}{$lm$$:\qty{22,21,33}$} \\
        \cmidrule(lr){2-5} \cmidrule(lr){7-10} \noalign{\vskip-9pt}
        \bf{event} & $\text{SNR}^{\text{net}}_{\text{opt}}$ & $H$ & $\text{lnL}_{\text{max}}$ & $\ln{\mathcal{B}_{lm,22}}$ & & $\text{SNR}^{\text{net}}_{\text{opt}}$ & $H$ & $\text{lnL}_{\text{max}}$ & $\ln{\mathcal{B}_{lm,22}}$    \\
        
        \midrule

        &&&&&&&&& \\[-6mm]
        \GW{150914}         & $12.5_{-1.3}^{+1.3}$ & $14.8$ & \scriptsize{$38775$} & $-0.43$ & & $12.5_{-1.3}^{+1.3}$ & $15.0$ & \scriptsize{$38775$} & $-0.70$ \\[-2mm]
        \GW{170729}         & $ 5.7_{-1.5}^{+1.4}$ & $ 9.5$ & \scriptsize{$57693$} & $ 0.93$ & & $ 5.7_{-1.5}^{+1.5}$ & $ 9.6$ & \scriptsize{$57694$} & $ 0.87$ \\[-2mm]
        \GW{170823}         & $ 6.2_{-1.4}^{+1.3}$ & $ 9.0$ & \scriptsize{$38952$} & $-0.15$ & & $ 6.1_{-1.4}^{+1.4}$ & $ 9.1$ & \scriptsize{$38952$} & $-0.29$ \\[-2mm]
        
        \GW{190503\_185404} & $ 5.7_{-1.4}^{+1.4}$ & $ 8.9$ & \scriptsize{$58528$} & $-0.32$ & & $ 5.7_{-1.4}^{+1.4}$ & $ 9.1$ & \scriptsize{$58528$} & $-0.61$ \\[-2mm]
        \GW{190519\_153544} & $ 9.5_{-1.3}^{+1.3}$ & $10.0$ & \scriptsize{$58604$} & $-0.78$ & & $ 9.5_{-1.3}^{+1.3}$ & $10.0$ & \scriptsize{$58604$} & $-0.76$ \\[-2mm]
        \GW{190521}         & $12.2_{-1.3}^{+1.3}$ & $10.9$ & \scriptsize{$58562$} & $-0.20$ & & $12.2_{-1.3}^{+1.3}$ & $11.1$ & \scriptsize{$58562$} & $-0.49$ \\[-2mm]
        \GW{190521\_074359} & $ 9.7_{-1.3}^{+1.3}$ & $12.8$ & \scriptsize{$39415$} & $ 0.14$ & & $ 9.7_{-1.3}^{+1.3}$ & $12.7$ & \scriptsize{$39415$} & $ 0.36$ \\[-2mm]
        \GW{190602\_175927} & $ 8.7_{-1.3}^{+1.3}$ & $ 9.8$ & \scriptsize{$58545$} & $-0.71$ & & $ 8.7_{-1.3}^{+1.3}$ & $ 9.9$ & \scriptsize{$58546$} & $-0.91$ \\[-2mm]
        \GW{190706\_222641} & $ 7.8_{-1.4}^{+1.3}$ & $11.4$ & \scriptsize{$58457$} & $-0.62$ & & $ 7.8_{-1.4}^{+1.4}$ & $11.4$ & \scriptsize{$58458$} & $-0.68$ \\[-2mm]
        \GW{190727\_060333} & $ 5.9_{-1.4}^{+1.4}$ & $ 8.5$ & \scriptsize{$58580$} & $ 0.03$ & & $ 5.9_{-1.4}^{+1.4}$ & $ 8.5$ & \scriptsize{$58581$} & $ 0.11$ \\[-2mm]
        \GW{190828\_063405} & $ 5.0_{-1.5}^{+1.4}$ & $ 9.1$ & \scriptsize{$58571$} & $-0.14$ & & $ 5.0_{-1.5}^{+1.4}$ & $ 9.1$ & \scriptsize{$58571$} & $-0.16$ \\[-2mm]
        \GW{190910\_112807} & $ 8.1_{-1.3}^{+1.3}$ & $10.8$ & \scriptsize{$38854$} & $-0.40$ & & $ 8.1_{-1.3}^{+1.3}$ & $10.9$ & \scriptsize{$38856$} & $-0.39$ \\[-2mm]
        \GW{190915\_235702} & $ 6.2_{-1.4}^{+1.3}$ & $10.6$ & \scriptsize{$58438$} & $-0.31$ & & $ 6.2_{-1.4}^{+1.4}$ & $10.6$ & \scriptsize{$58440$} & $-0.22$ \\[-2mm]
        
        \GW{191109\_010717} & $11.8_{-1.3}^{+1.3}$ & $15.1$ & \scriptsize{$39405$} & $-1.55$ & & $11.8_{-1.4}^{+1.3}$ & $15.2$ & \scriptsize{$39406$} & $-1.49$ \\[-2mm]
        \GW{191222\_033537} & $ 5.6_{-1.3}^{+1.3}$ & $ 7.4$ & \scriptsize{$39420$} & $-0.28$ & & $11.8_{-1.4}^{+1.3}$ & $15.2$ & \scriptsize{$39406$} & $-1.49$ \\[-2mm]
        \GW{200129\_065458} & $12.5_{-1.3}^{+1.3}$ & $14.0$ & \scriptsize{$58422$} & $ 0.59$ & & $12.5_{-1.3}^{+1.3}$ & $14.3$ & \scriptsize{$58422$} & $ 0.42$ \\[-2mm]
        \GW{200224\_222234} & $10.4_{-1.3}^{+1.3}$ & $13.0$ & \scriptsize{$58553$} & $ 0.44$ & & $10.4_{-1.3}^{+1.3}$ & $13.0$ & \scriptsize{$58553$} & $ 0.43$ \\[-2mm]
        \GW{200311\_115853} & $ 7.0_{-1.3}^{+1.3}$ & $11.4$ & \scriptsize{$58550$} & $-0.37$ & & $ 7.0_{-1.4}^{+1.4}$ & $11.5$ & \scriptsize{$58550$} & $-0.50$ \\[1mm]
        
        \bottomrule\bottomrule
    \end{tabular}
    
\caption{\footnotesize 
    Results from the HMs search.
    In the different columns, the optimal network signal-to-noise ratio $\text{SNR}^{\text{net}}_{\text{opt}}$, the information $H$, the logarithmic maximum likelihood $\text{lnL}_{\text{max}}$, and the logarithmic Bayes factor $\ln{\mathcal{B}_{lm,22}}$ between the model with HMs and the same with only the fundamental mode.
    These parameters are grouped by the different hypotheses considered: $\qty{(2,2)}$, $\qty{(2,2),(3,3)}$, $\qty{(2,2),(2,1)}$, $\qty{(2,2),(2,1),(3,3)}$.
    The reported values of $\text{SNR}^{\text{net}}_{\text{opt}}$ correspond to the median and $90\%$ symmetric credible intervals.}
    \label{tab:multimodal_analysis}
    
\end{table*}

% GW170729
\newpage
\begin{figure*}
    \includegraphics[scale=1.1]{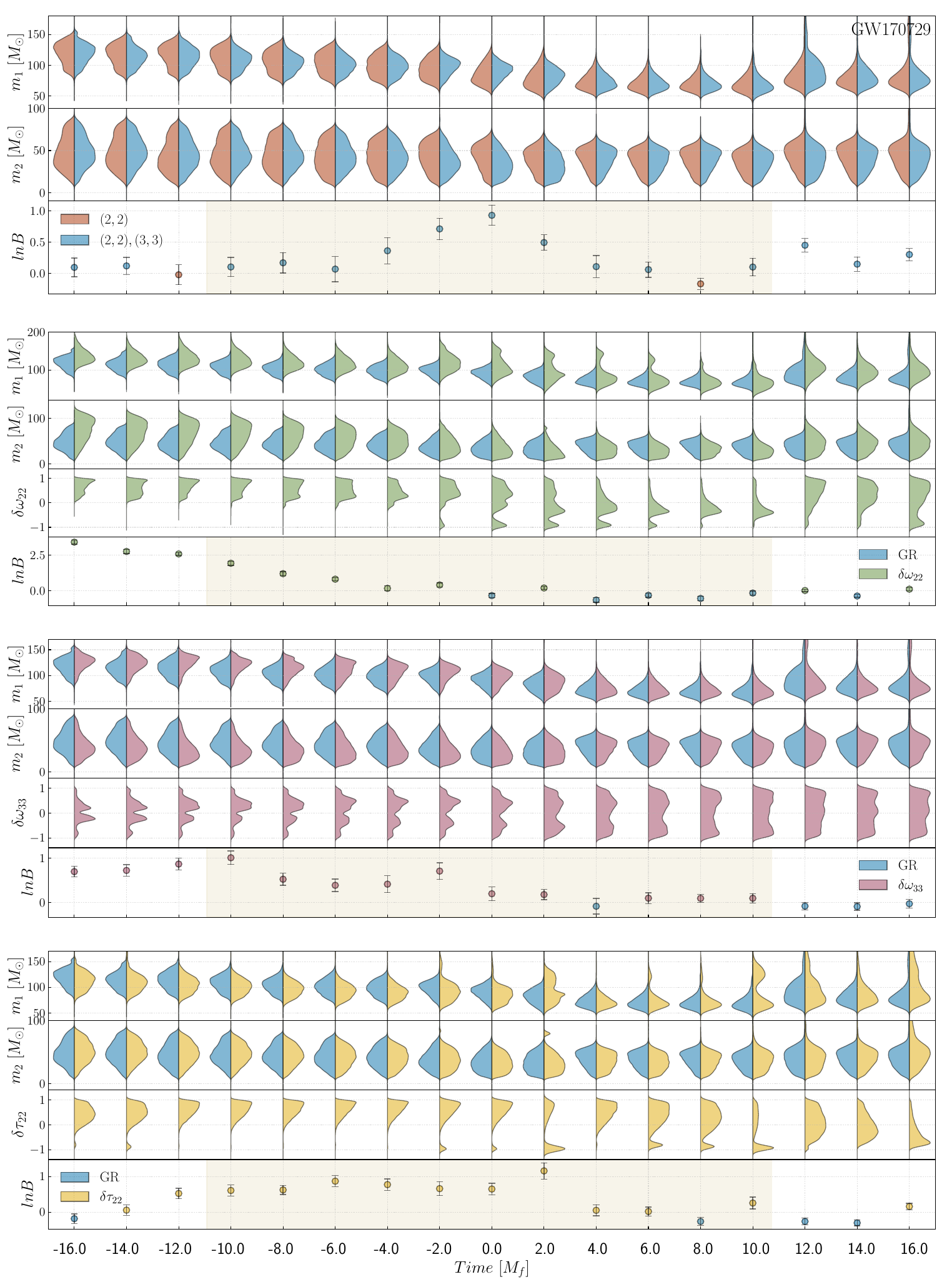}
    \caption{\footnotesize PE and model selection on the event \GW{170729}\footnotesize, as a function of the starting time.
    For each subplot, the first two rows display the initial masses $m_{1,2}$, and the bottom row the logarithmic Bayes factor between the competing hypotheses, with a color corresponding to the favoured hypothesis: \textit{red} and \textit{blue} assume GR with the modes $\qty{(2,2)}$ and $\qty{(2,2), (3,3)}$; \textit{green}, \textit{pink} and \textit{yellow} assume GR deviations on $\delta\omega_{22}$, $\delta\omega_{33}$ and $\delta\tau_{22}$.
    The shaded region in the bottom row outlines the $95\%$ CI of the corresponding IMR peaktime distribution.}
    \label{fig:GW170729}
\end{figure*}

% GW190521
\newpage
\begin{figure*}
    \includegraphics[scale=1.1]{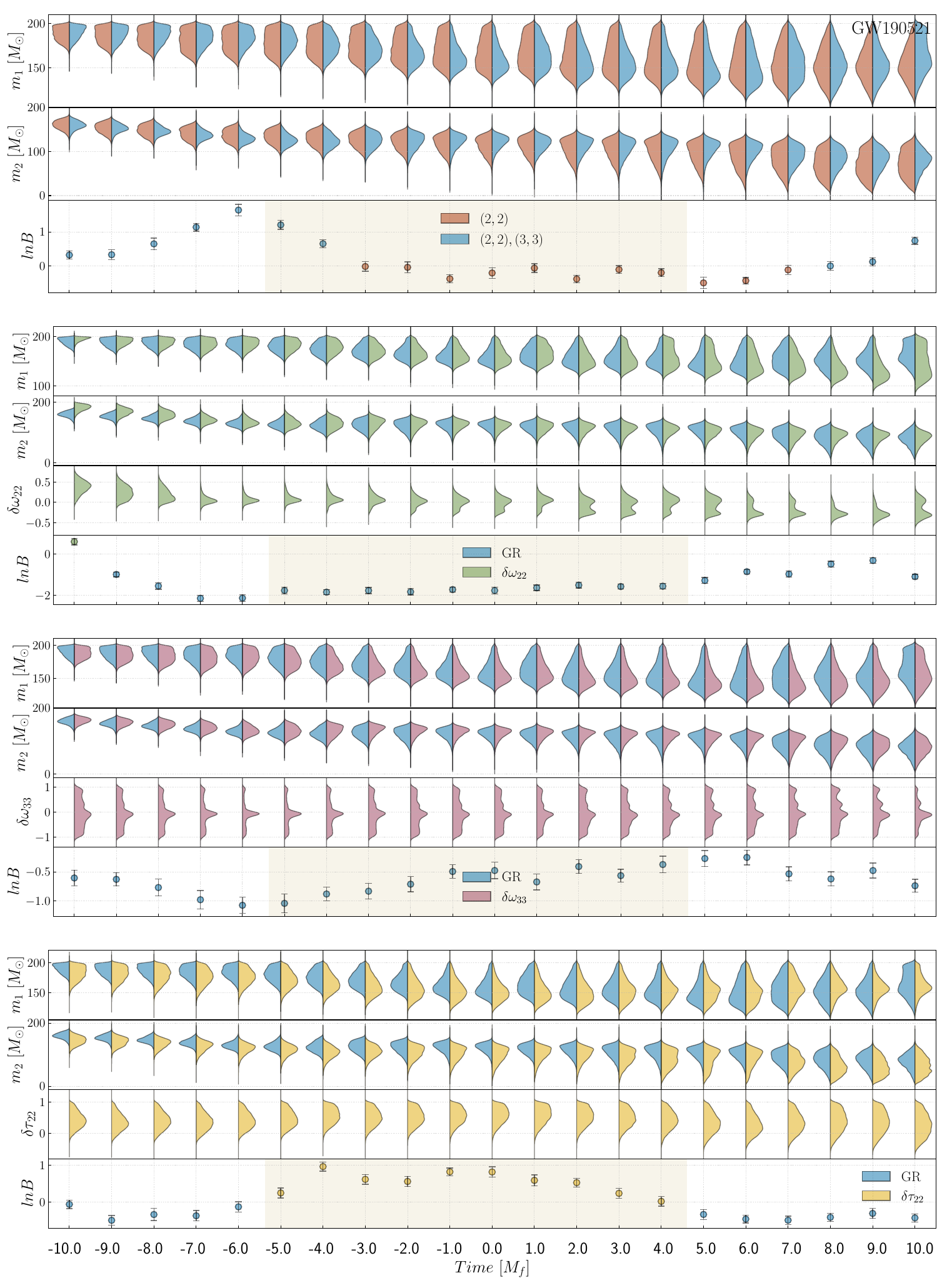}
    \caption{\footnotesize PE and model selection on the event \GW{190521}\footnotesize, as a function of the starting time.
    For each subplot, the first two rows display the initial masses $m_{1,2}$, and the bottom row the logarithmic Bayes factor between the competing hypotheses, with a color corresponding to the favoured hypothesis.: \textit{red} and \textit{blue} assume GR with the modes $\qty{(2,2)}$ and $\qty{(2,2), (3,3)}$; \textit{green}, \textit{pink} and \textit{yellow} assume GR deviations on $\delta\omega_{22}$, $\delta\omega_{33}$ and $\delta\tau_{22}$.
    The shaded region in the bottom row outlines the $95\%$ CI of the corresponding IMR peaktime distribution.}
    \label{fig:GW190521}
\end{figure*}

% GW191109_010717
\newpage
\begin{figure*}
    \includegraphics[scale=1.1]{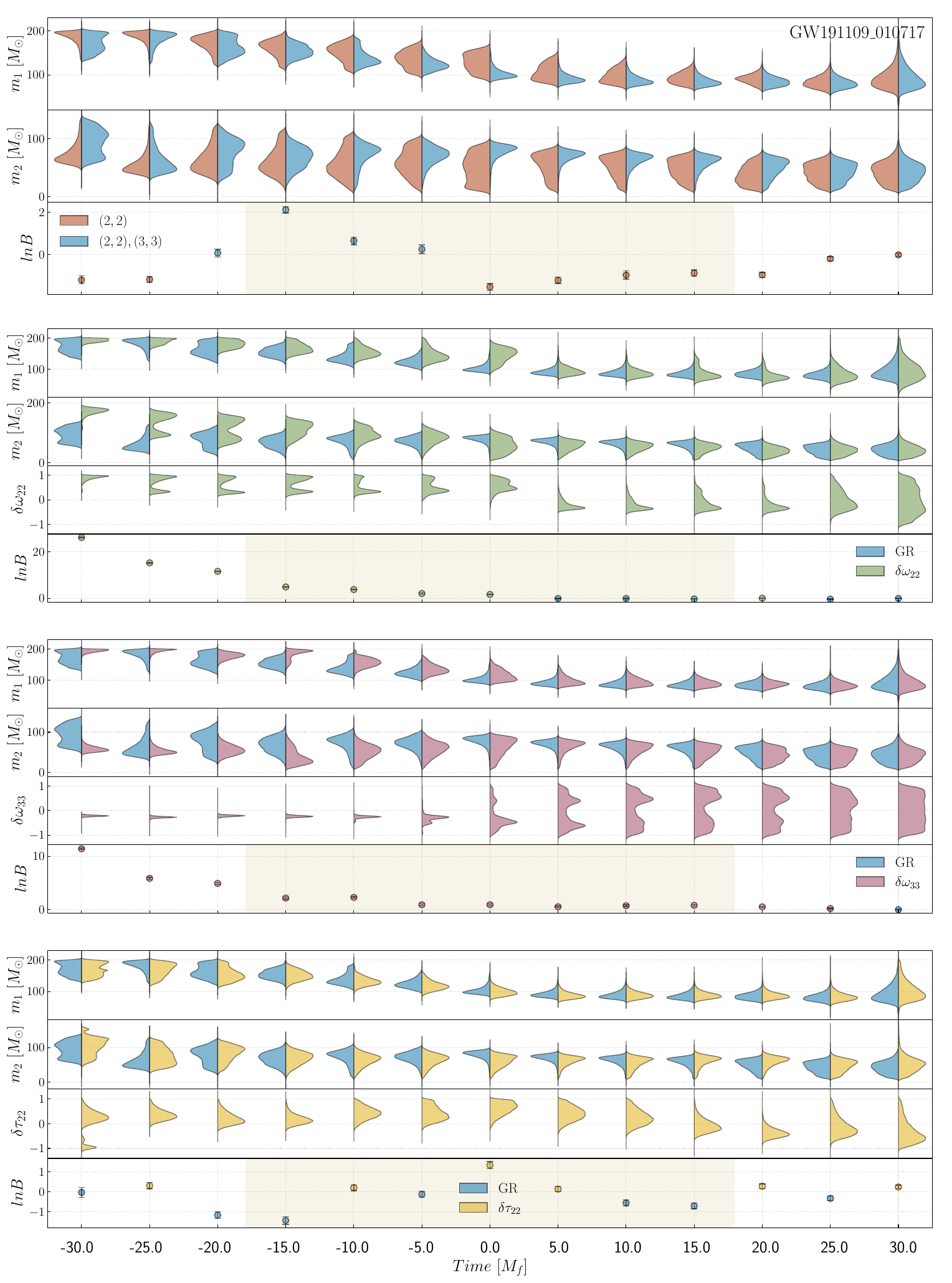}
    \caption{\footnotesize PE and model selection on the event \GW{191109\_010717}\footnotesize, as a function of the starting time.
    For each subplot, the first two rows display the initial masses $m_{1,2}$, and the bottom row the logarithmic Bayes factor between the competing hypotheses, with a color corresponding to the favoured hypothesis.: \textit{red} and \textit{blue} assume GR with the modes $\qty{(2,2)}$ and $\qty{(2,2), (3,3)}$; \textit{green}, \textit{pink} and \textit{yellow} assume GR deviations on $\delta\omega_{22}$, $\delta\omega_{33}$ and $\delta\tau_{22}$.
    The shaded region in the bottom row outlines the $95\%$ CI of the corresponding IMR peaktime distribution.}
    \label{fig:GW191109_010717}
\end{figure*}

% GW200129_065458
\newpage
\begin{figure*}
    \includegraphics[scale=1.1]{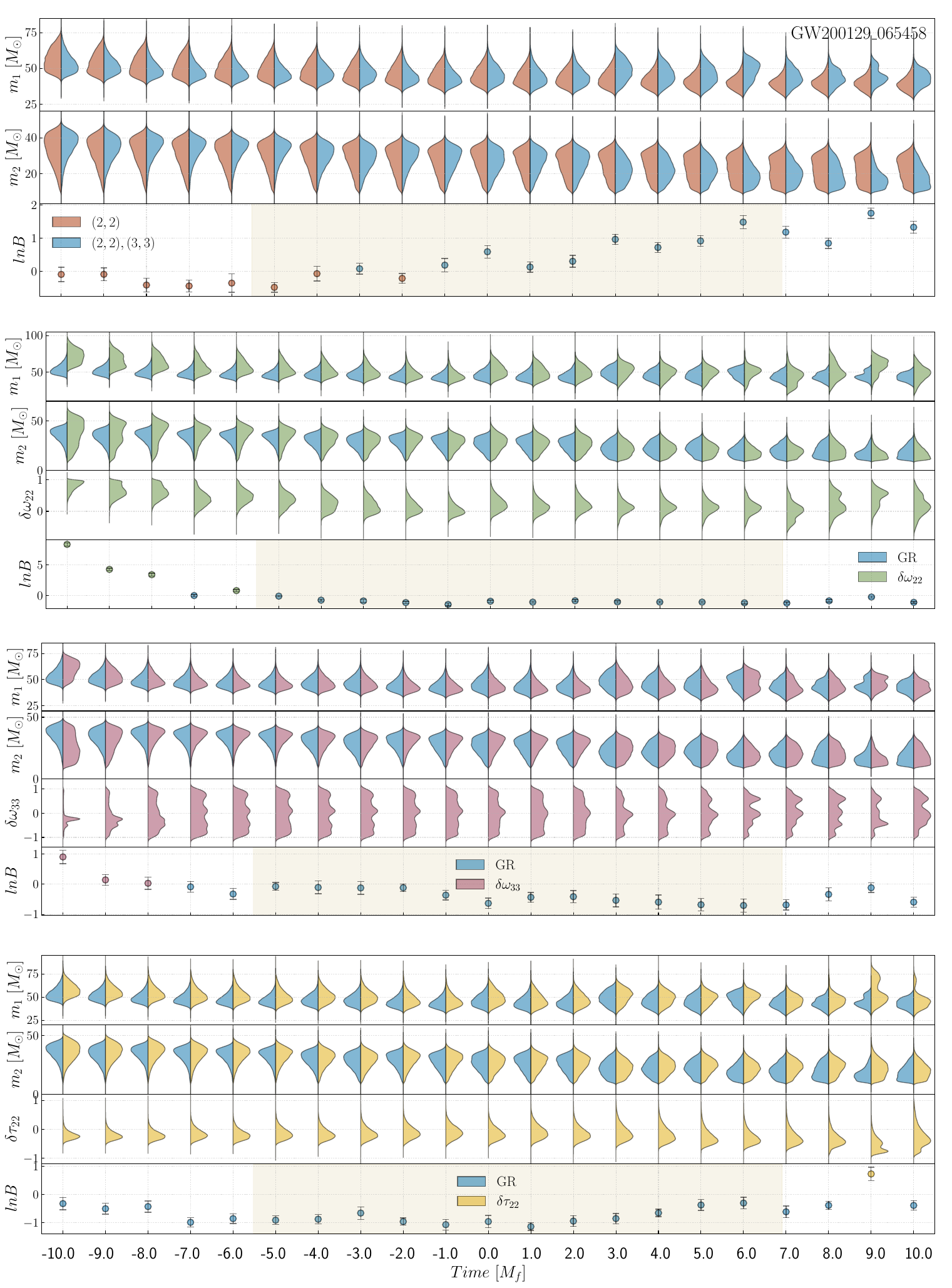}
    \caption{\footnotesize PE and model selection on the event \GW{200129\_065458}\footnotesize, as a function of the starting time.
    For each subplot, the first two rows display the initial masses $m_{1,2}$, and the bottom row the logarithmic Bayes factor between the competing hypotheses, with a color corresponding to the favoured hypothesis.: \textit{red} and \textit{blue} assume GR with the modes $\qty{(2,2)}$ and $\qty{(2,2), (3,3)}$; \textit{green}, \textit{pink} and \textit{yellow} assume GR deviations on $\delta\omega_{22}$, $\delta\omega_{33}$ and $\delta\tau_{22}$.
    The shaded region in the bottom row outlines the $95\%$ CI of the corresponding IMR peaktime distribution.}
    \label{fig:GW200129_065458}
\end{figure*}

% GW200224_222234
\newpage
\begin{figure*}
    \includegraphics[scale=1.1]{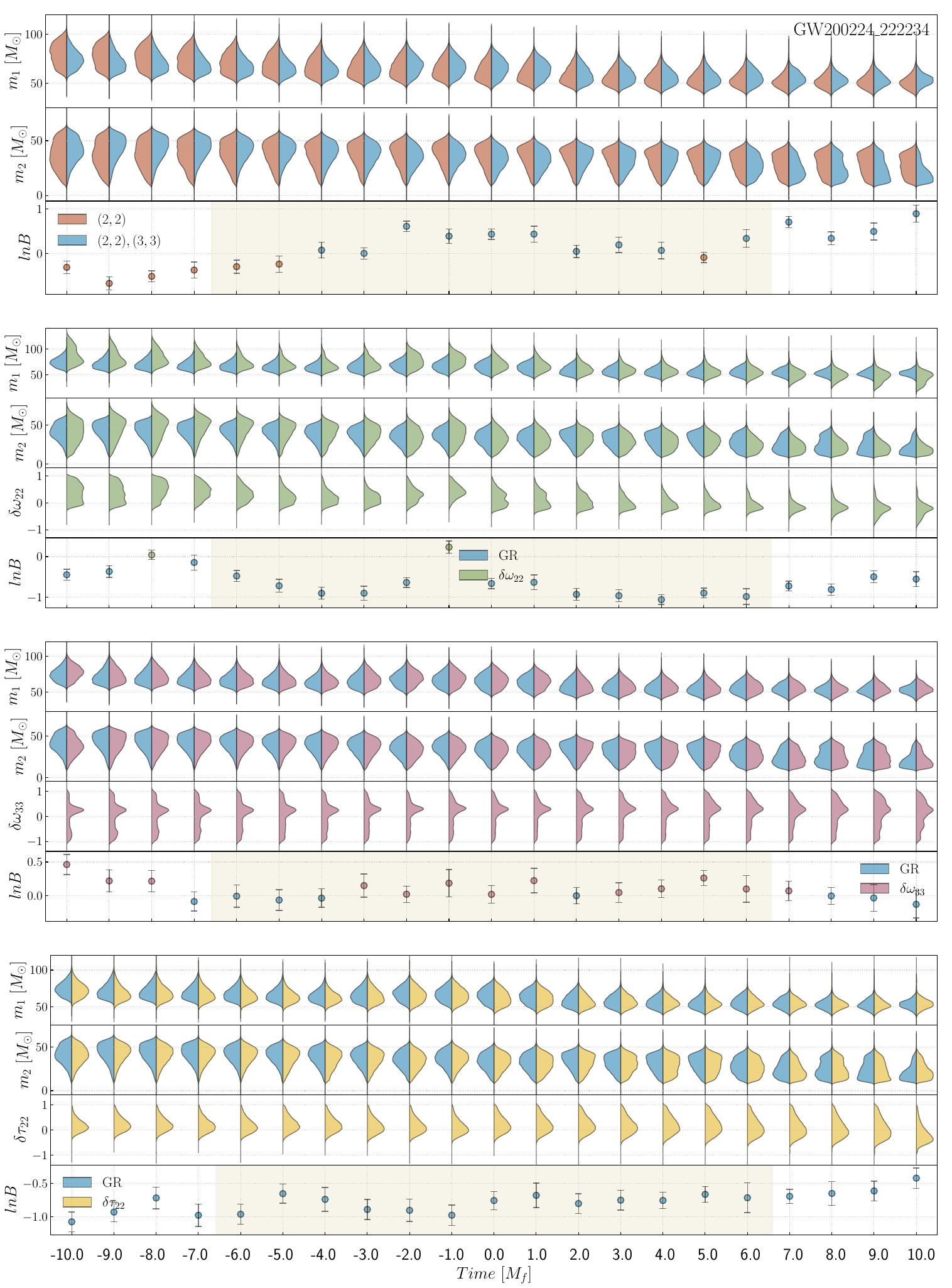}
    \caption{\footnotesize PE and model selection on the event \GW{200224\_222234}\footnotesize, as a function of the starting time.
    For each subplot, the first two rows display the initial masses $m_{1,2}$, and the bottom row the logarithmic Bayes factor between the competing hypotheses, with a color corresponding to the favoured hypothesis.: \textit{red} and \textit{blue} assume GR with the modes $\qty{(2,2)}$ and $\qty{(2,2), (3,3)}$; \textit{green}, \textit{pink} and \textit{yellow} assume GR deviations on $\delta\omega_{22}$, $\delta\omega_{33}$ and $\delta\tau_{22}$.
    The shaded region in the bottom row outlines the $95\%$ CI of the corresponding IMR peaktime distribution.}
    \label{fig:GW200224_222234}
\end{figure*}

% GW150914
\begin{figure*}
    \includegraphics[scale=1.1]{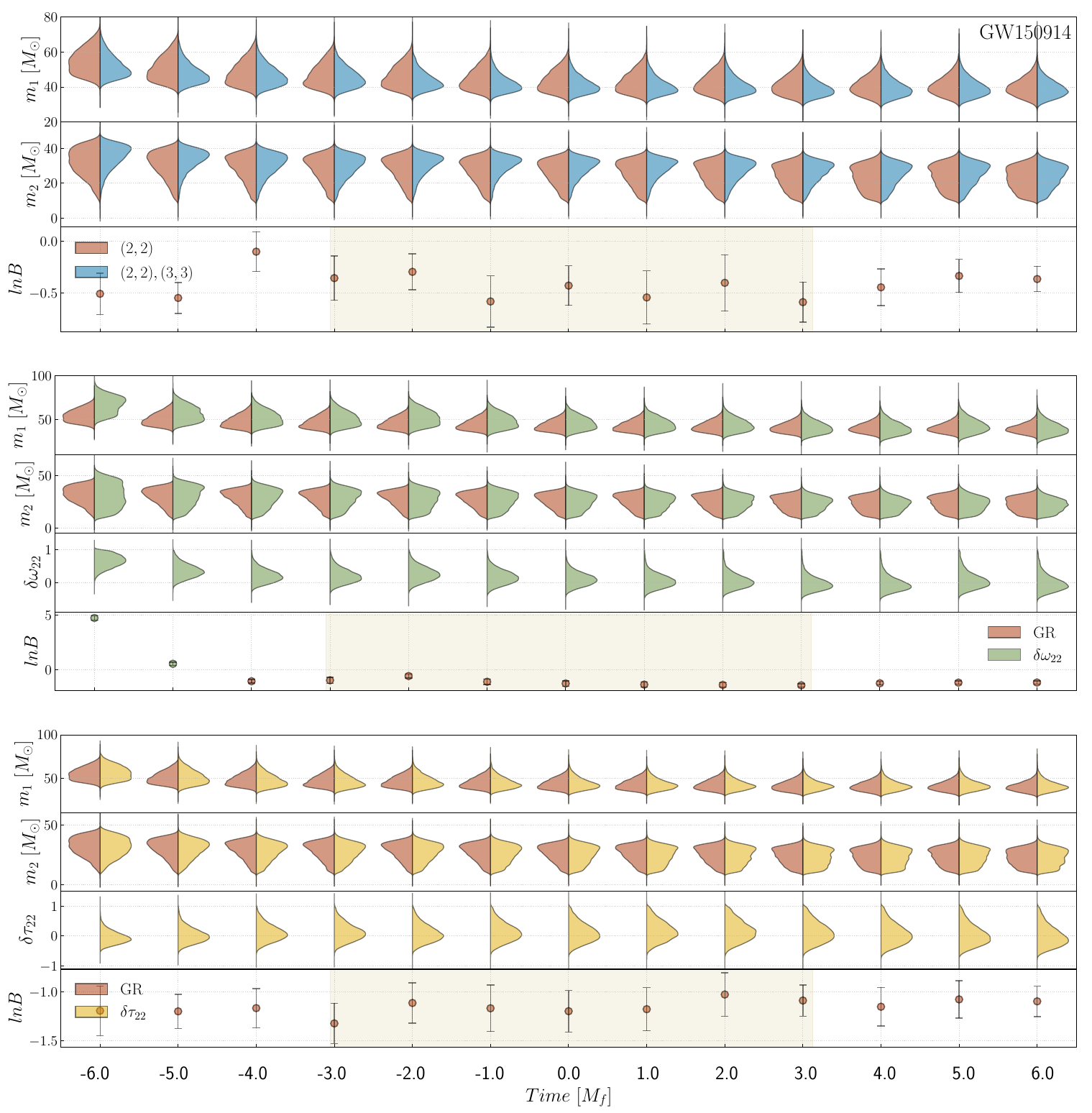}
    \caption{\footnotesize PE and model selection on the event \GW{150914}\footnotesize, as a function of the starting time.
    For each subplot, the first two rows display the initial masses $m_{1,2}$, and the bottom row the logarithmic Bayes factor between the competing hypotheses, with a color corresponding to the favoured hypothesis.: \textit{red} and \textit{blue} assume GR with the modes $\qty{(2,2)}$ and $\qty{(2,2), (3,3)}$; \textit{green} and \textit{yellow} assume GR deviations on $\delta\omega_{22}$ and $\delta\tau_{22}$.
    The shaded region in the bottom row outlines the $95\%$ CI of the corresponding IMR peaktime distribution.}
    \label{fig:GW150914}
\end{figure*}

% TGR fundamental mode
\begin{figure*}
    \includegraphics[width=\textwidth, scale=1.1]{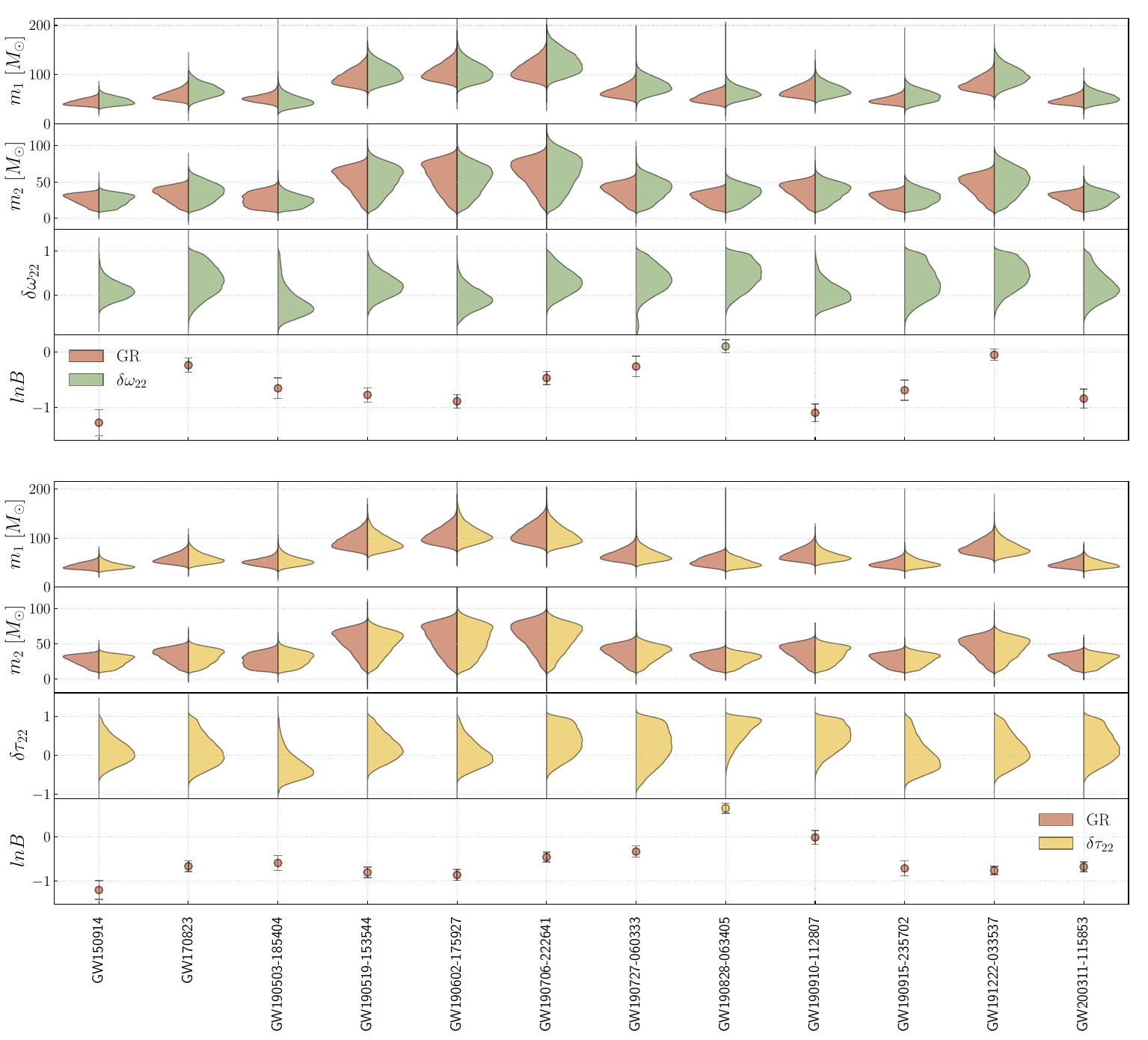}
    \caption{\footnotesize Posteriors distributions of the initial masses $m_{1,2}$ and deviations from GR, for the different events on the x-axis.
    The \textit{red} posteriors are from the GR analysis with only the fundamental mode $(2,2)$, the \textit{green} ones include fractional deviations on the frequency of the fundamental mode $\delta\omega_{22}$, the \textit{yellow} ones include fractional deviations on the damping time of the fundamental mode $\delta\tau_{22}$.
    For the two subplots, the bottom row shows the logarithmic Bayes factor between the GR and non-GR analysis, for the competing hypothesis in colors.}
    \label{fig:TGR_fundamental}
\end{figure*}

% TEOBPM mismatches
\newpage
\begin{figure*}
    \begin{center}
        \hspace*{-0.6cm}\includegraphics[scale=0.92]{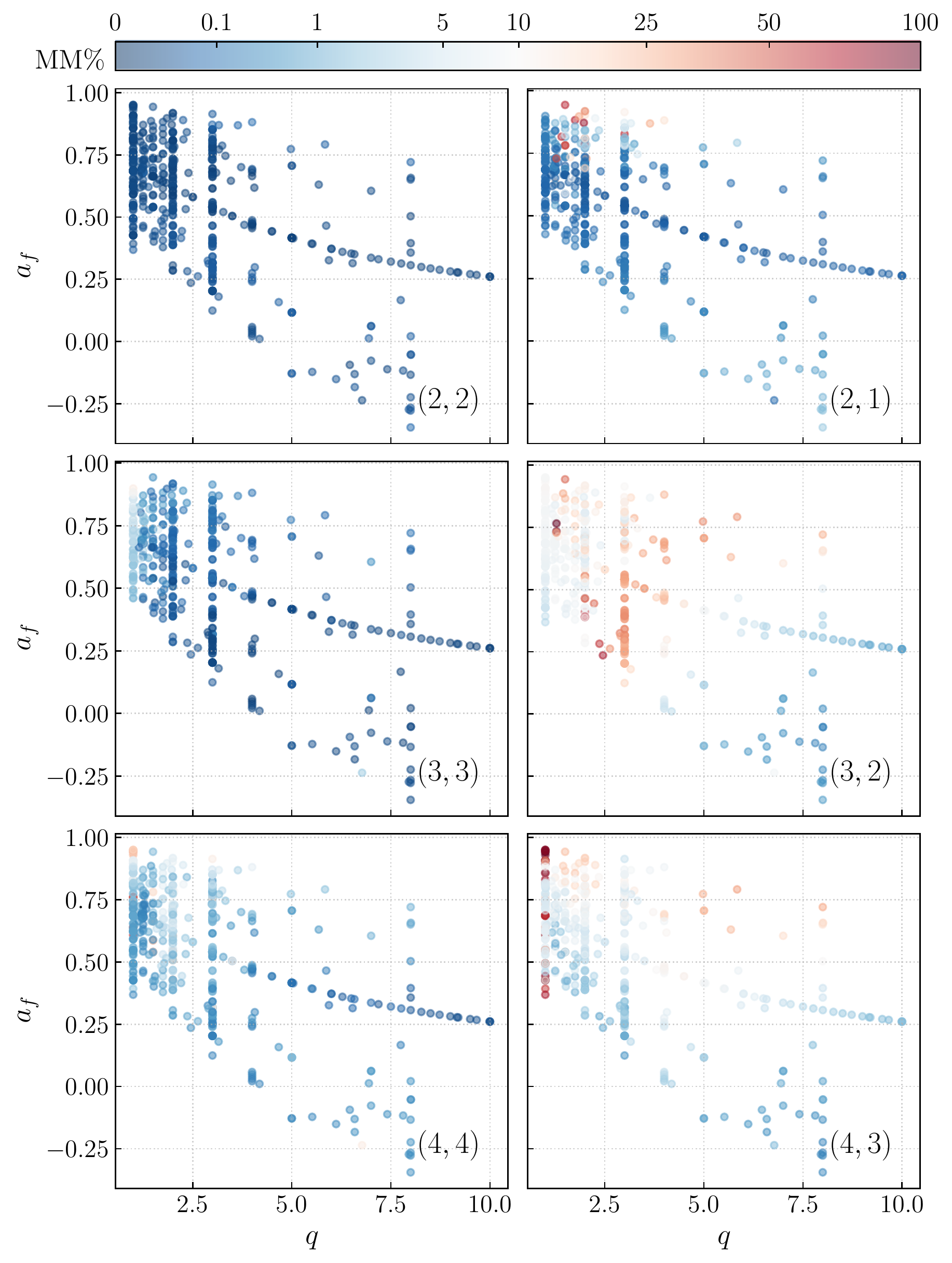}
        \caption{\footnotesize Comparison between the numerical simulation from the SXS catalog and the \texttt{TEOBPM} implementation in \texttt{pyRing}.
        For the modes $\qty{(2,2),(2,1),(3,3),(3,2),(4,4),(4,3)}$, in colours is represented the percent mismatch $\text{MM}\%$ of the quantities $h_{lm}$ in Eq.~\eqref{TEOBPM_multimode_template}, as a function of the mass ratio $q$ and adimensional spin of the final BH $a_f$.
        Each point in the plot corresponds to a SXS simulation.}
        \label{mismatch_catalog_h_fig}
    \end{center}
\end{figure*}

% Table pyRing inputs
\begin{table*}
\setlength{\tabcolsep}{8pt}
\scalebox{1.1}{
\begin{tabular}{ lllll }
    \hline
    \hline
    event & $t_0$ & $\alpha$ & $\delta$ & IFOs \\
    & $[\text{GPS}]$ & $[\text{rad}]$ & $[\text{rad}]$ & \\
    \hline
    
    \GW{150914}         & \footnotesize{$1126259462.4232266$} & \footnotesize{$1.1578762186634035$}  & \footnotesize{$-1.191080211569962$}   & \small{HL}  \\[0.3mm]
    \GW{151012}         & \footnotesize{$1128678900.4420469$} & \footnotesize{$5.344926802197244$}   & \footnotesize{$-1.019757877894778$}   & \small{HL}  \\[0.3mm]
    \GW{151226}         & \footnotesize{$1135136350.648$}     & \footnotesize{$3.445889773508687$}   & \footnotesize{$-0.10661115666629918$} & \small{HL}  \\[0.3mm]
    \GW{170104}         & \footnotesize{$1167559936.593629$}  & \footnotesize{$2.0081441006460006$}  & \footnotesize{$ 0.14416278221335294$} & \small{HL}  \\[0.3mm]
    \GW{170608}         & \footnotesize{$1180922494.4836993$} & \footnotesize{$2.0892021117657333$}  & \footnotesize{$ 0.3567599522292868$}  & \small{HL}  \\[0.3mm]
    \GW{170729}         & \footnotesize{$1185389807.3138866$} & \footnotesize{$5.704589981121234$}   & \footnotesize{$-1.194842248042486$}   & \small{HLV} \\[0.3mm]
    \GW{170809}         & \footnotesize{$1186302519.7494414$} & \footnotesize{$0.2584735357365375$}  & \footnotesize{$-0.507529253902754$}   & \small{HLV} \\[0.3mm]
    \GW{170814}         & \footnotesize{$1186741861.5307062$} & \footnotesize{$0.8083063830865238$}  & \footnotesize{$-0.7970854269814471$}  & \small{HLV} \\[0.3mm]
    \GW{170818}         & \footnotesize{$1187058327.0814648$} & \footnotesize{$5.9545154473923905$}  & \footnotesize{$ 0.37181209700816104$} & \small{HLV} \\[0.3mm]
    \GW{170823}         & \footnotesize{$1187529256.5101874$} & \footnotesize{$4.648229200129752$}   & \footnotesize{$-0.7777112385699141$}  & \small{HL}  \\
    [3mm]

    \GW{190408\_181802} & \footnotesize{$1238782700.2813442$} & \footnotesize{$0.19422991378016619$} & \footnotesize{$ 1.1992821930577033$}  & \small{HLV} \\[0.3mm]
    \GW{190412}         & \footnotesize{$1239082262.1645508$} & \footnotesize{$3.819586104366042$}   & \footnotesize{$ 0.6274962671486665$}  & \small{HLV} \\[0.3mm]
    \GW{190421\_213856} & \footnotesize{$1239917954.250977$}  & \footnotesize{$3.475941261508279$}   & \footnotesize{$-0.4996728211179956$}  & \small{HL}  \\[0.3mm]
    \GW{190503\_185404} & \footnotesize{$1240944862.2912657$} & \footnotesize{$1.6530573709519114$}  & \footnotesize{$-0.8710969233362608$}  & \small{HLV} \\[0.3mm]
    \GW{190512\_180714} & \footnotesize{$1241719652.4145505$} & \footnotesize{$4.262454975648534$}   & \footnotesize{$-0.39930209189187016$} & \small{HLV} \\[0.3mm]
    \GW{190513\_205428} & \footnotesize{$1241816086.7412295$} & \footnotesize{$4.973410757909485$}   & \footnotesize{$-0.5083449594497929$}  & \small{HLV} \\[0.3mm]
    \GW{190517\_055101} & \footnotesize{$1242107479.8257282$} & \footnotesize{$4.003025310084514$}   & \footnotesize{$-0.7313982894457726$}  & \small{HLV} \\[0.3mm]
    \GW{190519\_153544} & \footnotesize{$1242315362.3788562$} & \footnotesize{$3.2485874621900575$}  & \footnotesize{$-0.3898261701379244$}  & \small{HLV} \\[0.3mm]
    \GW{190521}         & \footnotesize{$1242442967.4287114$} & \footnotesize{$0.16416072898840461$} & \footnotesize{$-1.1434284511661967$}  & \small{HLV} \\[0.3mm]
    \GW{190521\_074359} & \footnotesize{$1242459857.4604406$} & \footnotesize{$5.455734590393311$}   & \footnotesize{$ 0.5865436466945467$}  & \small{HL}  \\[0.3mm]
    \GW{190602\_175927} & \footnotesize{$1243533585.0854938$} & \footnotesize{$1.3334423596499836$}  & \footnotesize{$-0.7553909477951349$}  & \small{HLV} \\[0.3mm]
    \GW{190630\_185205} & \footnotesize{$1245955943.1664371$} & \footnotesize{$5.935798679516984$}   & \footnotesize{$-0.13643390906247874$} & \small{LV}  \\[0.3mm]
    \GW{190706\_222641} & \footnotesize{$1246487219.3173833$} & \footnotesize{$1.4563931181833372$}  & \footnotesize{$-0.9090358111010055$}  & \small{HLV} \\[0.3mm]
    \GW{190707\_093326} & \footnotesize{$1246527224.1675541$} & \footnotesize{$4.98170434655823$}    & \footnotesize{$-0.15890694388859383$} & \small{HL}  \\[0.3mm]
    \GW{190708\_232457} & \footnotesize{$1246663515.3815293$} & \footnotesize{$3.602711227860004$}   & \footnotesize{$ 1.0878104409201002$}  & \small{LV}  \\[0.3mm]
    \GW{190720\_000836} & \footnotesize{$1247616534.7045903$} & \footnotesize{$5.192207387604943$}   & \footnotesize{$ 0.6536557311651274$}  & \small{HLV} \\[0.3mm]
    \GW{190727\_060333} & \footnotesize{$1248242631.9800277$} & \footnotesize{$2.499642949119032$}   & \footnotesize{$-1.21678298573251$}    & \small{HLV} \\[0.3mm]
    \GW{190728\_064510} & \footnotesize{$1248331528.532623$}  & \footnotesize{$5.486464527281894$}   & \footnotesize{$ 0.17580234224696964$} & \small{HLV} \\[0.3mm]
    \GW{190814}         & \footnotesize{$1249852257.0068965$} & \footnotesize{$0.21787450687454352$} & \footnotesize{$-0.43496688165485414$} & \small{HLV} \\[0.3mm]
    \GW{190828\_063405} & \footnotesize{$1251009263.7515893$} & \footnotesize{$2.4534823079417425$}  & \footnotesize{$-0.40402553657088913$} & \small{HLV} \\[0.3mm]
    \GW{190828\_065509} & \footnotesize{$1251010527.8850923$} & \footnotesize{$2.265508889735858$}   & \footnotesize{$-0.4706874299709763$}  & \small{HLV} \\[0.3mm]
    \GW{190910\_112807} & \footnotesize{$1252150105.3154283$} & \footnotesize{$1.53682178616314$}    & \footnotesize{$ 1.1610432574944787$}  & \small{LV}  \\[0.3mm]
    \GW{190915\_235702} & \footnotesize{$1252627040.6850317$} & \footnotesize{$3.360607999494463$}   & \footnotesize{$ 0.17994778522861346$} & \small{HLV} \\[0.3mm]
    \GW{190924\_021846} & \footnotesize{$1253326744.839818$}  & \footnotesize{$2.191336654308782$}   & \footnotesize{$ 0.3631486729464257$}  & \small{HLV} \\
    [3mm]
    
    \GW{191109\_010717} & \footnotesize{$1257296855.1979034$} & \footnotesize{$3.730943133832227$}   & \footnotesize{$-0.630875819578621$}   & \small{HL}  \\[0.3mm]
    \GW{191129\_134029} & \footnotesize{$1259070047.198897$}  & \footnotesize{$5.692492322954507$}   & \footnotesize{$-0.6895738841121974$}  & \small{HL}  \\[0.3mm]
    \GW{191204\_171526} & \footnotesize{$1259514944.0872023$} & \footnotesize{$1.877108449500568$}   & \footnotesize{$-0.735532575154993$}   & \small{HL}  \\[0.3mm]
    \GW{191215\_223052} & \footnotesize{$1260484270.3314903$} & \footnotesize{$5.7050225595398345$}  & \footnotesize{$ 0.3487593151779902$}  & \small{HLV} \\[0.3mm]
    \GW{191216\_213338} & \footnotesize{$1260567236.4708745$} & \footnotesize{$5.509998419175801$}   & \footnotesize{$ 0.5720639112948416$}  & \small{HV}  \\[0.3mm]
    \GW{191222\_033537} & \footnotesize{$1261020955.114826$}  & \footnotesize{$3.81569413468133$}    & \footnotesize{$-0.8647541697031778$}  & \small{HL}  \\[0.3mm]
    \GW{200129\_065458} & \footnotesize{$1264316116.4219835$} & \footnotesize{$5.553763898571937$}   & \footnotesize{$ 0.09477498129492656$} & \small{HLV} \\[0.3mm]
    \GW{200208\_130117} & \footnotesize{$1265202095.940918$}  & \footnotesize{$2.432916400916298$}   & \footnotesize{$-0.6008586743005835$}  & \small{HLV} \\[0.3mm]
    \GW{200219\_065458} & \footnotesize{$1266140673.189855$}  & \footnotesize{$0.3150678803598552$}  & \footnotesize{$-0.4247639842579993$}  & \small{HLV} \\[0.3mm]
    \GW{200202\_154313} & \footnotesize{$1264693411.5546873$} & \footnotesize{$2.5517572415865715$}  & \footnotesize{$ 0.3275445136801054$}  & \small{HLV} \\[0.3mm]
    \GW{200224\_222234} & \footnotesize{$1266618172.3968902$} & \footnotesize{$3.050654724133011$}   & \footnotesize{$-0.16712575394750373$} & \small{HLV} \\[0.3mm]
    \GW{200225\_060421} & \footnotesize{$1266645879.3935542$} & \footnotesize{$1.795645173012197$}   & \footnotesize{$ 0.6037234211630302$}  & \small{HL}  \\[0.3mm]
    \GW{200311\_115853} & \footnotesize{$1267963151.3921406$} & \footnotesize{$0.04880293379906675$} & \footnotesize{$-0.16128963890506864$} & \small{HLV} \\[0.3mm]
    \GW{200316\_215756} & \footnotesize{$1268431094.1580691$} & \footnotesize{$1.5019364419658956$}  & \footnotesize{$ 0.8429868038645901$}  & \small{HLV} \\[0.3mm]

    \hline
    \hline
    \end{tabular}
    }
    \caption{\footnotesize Input parameters used in our analysis from IMR for each event analysed.
    $t_0$ being the starting time of the analysis in GPS time, $\qty{\alpha, \delta}$ being the right ascension and declination.
    The last column indicates the interferometers (IFOs) used for that event.
    }
    \label{tab:input_parameters_pyring_table}
\end{table*}

%%%%%%%%%%%%%%%%%%%%%%%%%%%%%%%%%%%%%%%%%%%%%%%%%%%%%%%%%%%%%%%%%%%%%%%%%%%%%%%
\clearpage
\bibliography{references}

%%%%%%%%%%%%%%%%%%%%%%%%%%%%%%%%%%%%%%%%%%%%%%%%%%%%%%%%%%%%%%%%%%%%%%%%%%%%%%%
\end{document}